\pdfoutput=1

\documentclass[preprint]{acmart}

\usepackage{microtype}
\usepackage{booktabs} % For formal tables

\usepackage{amssymb,amsmath,latexsym,amsthm,stmaryrd}
\usepackage{csquotes}

\usepackage{tikz}
\usetikzlibrary{decorations.pathmorphing,shapes,arrows}
\usepackage{listings}
\usepackage{pifont}

\lstdefinestyle{gcore}{
  belowcaptionskip=1\baselineskip,
  breaklines=true,
  xleftmargin=\parindent,
  language=SQL,
  numbers=left,
  firstnumber=last,
  numbersep=5pt,
  showstringspaces=false,
  basicstyle=\footnotesize\ttfamily,
  keywordstyle=\bfseries\color{green!40!black},
  commentstyle=\itshape\color{purple!40!black},
  identifierstyle=\color{blue},
  stringstyle=\color{black},
  morekeywords={COST,PATH,CONSTRUCT,SHORTEST,OPTIONAL,REMOVE,IS,GRAPH,SUBSET}
}
\setcopyright{rightsretained}
\newcommand{\com}[2]{{\color{red} {\bf #1: #2}}}
\newcommand{\gcore}{\text{\sc G-CORE}}
\newcommand{\gcgraph}{\text{\sc PPG}}
\newcommand{\bL}{\text{\bf L}}
\newcommand{\bK}{\text{\bf K}}
\newcommand{\bV}{\text{\bf V}}
\newcommand{\cV}{\mathcal{V}}
\newcommand{\cN}{\mathcal{N}}
\newcommand{\cE}{\mathcal{E}}
\newcommand{\cP}{\mathcal{P}}
\newcommand{\fset}{\text{FSET}}
\newcommand{\flist}{\text{FLIST}}

\newcommand{\ignore}[1]{}

\newcommand{\nodes}[1]{\ensuremath{\op{nodes}(#1)}}
\newcommand{\edges}[1]{\ensuremath{\op{edges}(#1)}}

\newcommand{\noop}[1]{}

\newcommand{\set}[1]{\ensuremath{\left\{#1\right\}}} % a set; \set{X} gives {X}
\newcommand{\brk}[1]{\ensuremath{\left(#1\right)}} % a record; \brk{X} gives (X)
\newcommand{\op}[1]{\ensuremath{\texttt{#1}}} % an operation; \op{X} gives operation X in typewriter type face
\newcommand{\dom}[1]{\ensuremath{\text{dom}(#1)}}

\newcommand{\elt}[1]{\ensuremath{\textsf{#1}}}

\makeatletter
\newcommand\xleftrightarrow[2][]{%
  \ext@arrow 9999{\longleftrightarrowfill@}{#1}{#2}}
\newcommand\longleftrightarrowfill@{%
  \arrowfill@\leftarrow\relbar\rightarrow}
\makeatother

\newcommand{\edgepattern}[3]{\ensuremath{#1\xrightarrow{#2}#3}}

\newcounter{sarrow}
\newcommand\xrsquigarrow[1]{%
\stepcounter{sarrow}%
\begin{tikzpicture}[decoration={snake,segment length=1.5mm,amplitude=0.5mm}]
\node[inner ysep=1pt] (\thesarrow) {\strut\ensuremath{\scriptstyle#1}};
\draw[-implies,double equal sign distance] (\thesarrow.south west) -- (\thesarrow.south east);
\end{tikzpicture}%
}

\newcommand{\pathpattern}[4]{\ensuremath{#1\;\xrsquigarrow{@#2\text{ in }#3}\;#4}}

\newcommand{\wpathpattern}[4]{\ensuremath{#1\;\xrsquigarrow{#2\text{ in }#3}\;#4}}

\newcommand{\eval}[2]{\ensuremath{\llbracket #1 \rrbracket_{#2}}}

\newcommand{\und}{\underline{\phantom{a}}}

\def\ojoin{\setbox0=\hbox{$\Join$}%
\rule[0.11ex]{.28em}{.4pt}\llap{\rule[1.39ex]{.28em}{.4pt}}}
\def\leftouterjoin{\mathbin{\Join\mkern-6.4mu\ojoin}}

\newcommand{\kw}[1]{\ensuremath{\text{\sc #1}}}
\newcommand{\match}{\kw{Match}}
\newcommand{\optional}{\kw{Optional}}
\newcommand{\where}{\kw{Where}}
\newcommand{\construct}{\kw{Construct}}
\newcommand{\setcl}{\kw{Set}}
\newcommand{\remove}{\kw{Remove}}
\newcommand{\when}{\kw{When}}
\newcommand{\Path}{\kw{Path}}

\newcommand{\cost}{\kw{Cost}}
\newcommand{\Group}{\kw{Group}}
\newcommand{\Union}{\kw{Union}}
\newcommand{\Intersect}{\kw{Intersect}}
\newcommand{\Minus}{\kw{Minus}}
\newcommand{\Exists}{\kw{Exists}}
\newcommand{\Not}{\kw{Not}}
\newcommand{\Andkw}{\kw{And}}
\newcommand{\Orkw}{\kw{Or}}
\newcommand{\Subsetof}{\kw{Subset Of}}
\newcommand{\In}{\kw{In}}
\newcommand{\Labels}{\kw{Labels}}
\newcommand{\Nodes}{\kw{Nodes}}
\newcommand{\Edges}{\kw{Edges}}
\newcommand{\Size}{\kw{Size}}
\newcommand{\Count}{\kw{Count}}
\newcommand{\Min}{\kw{Min}}
\newcommand{\Max}{\kw{Max}}
\newcommand{\Avg}{\kw{Avg}}
\newcommand{\Sum}{\kw{Sum}}
\newcommand{\Collect}{\kw{Collect}}
\newcommand{\Graph}{\kw{Graph}}
\newcommand{\View}{\kw{View}}

\newcommand{\mw}[2]{\ensuremath{\match \ #1 \ \where \ #2}}
\newcommand{\ow}[2]{\ensuremath{\optional \ #1 \ \where \ #2}}
\newcommand{\csr}[1]{\ensuremath{\construct \ #1 }}%\ \setcl \ #2 \ \remove \ #3}}

\newcommand{\nl}[1]{\text{!}#1}

\newcommand{\lbgp}{\text{\it walkPattern}}
\newcommand{\bgps}{\text{\it graphPattern}}
\newcommand{\bgp}{\text{\it basicGraphPattern}}
\newcommand{\np}{\text{\it nodePattern}}
\newcommand{\ep}{\text{\it edgePattern}}
\newcommand{\pp}{\text{\it pathPattern}}
\newcommand{\fgp}{\text{\it fullGraphPattern}}

\newcommand{\ppattern}{\text{\it pathPattern}}

\newcommand{\query}{\text{\it query}}

\newcommand{\fgquery}{\text{\it fullGraphQuery}}
\newcommand{\bgquery}{\text{\it basicGraphQuery}}

\newcommand{\hclause}{\text{\it headClause}}
\newcommand{\pclause}{\text{\it pathClause}}
\newcommand{\bpclause}{\text{\it basicPathClause}}
\newcommand{\pwcclause}{\text{\it basicPathClauseCost}}

\newcommand{\gclause}{\text{\it graphClause}}
\newcommand{\gv}{\text{\it graphView}}
\newcommand{\cclause}{\text{\it constructClause}}
\newcommand{\mclause}{\text{\it matchClause}}
\newcommand{\fgpc}{\text{\it fullGraphPatternCondition}}
\newcommand{\oclause}{\text{\it optionalClause}}
\newcommand{\bc}{\text{\it BooleanCondition}}
\newcommand{\bgpl}{\text{\it basicGraphPatternLocation}}
\newcommand{\sopt}{\text{\it setOp}}
\newcommand{\loc}{\text{\it location}}
\newcommand{\on}{\text{\sc On}}
\newcommand{\gid}{\text{\it gid}}
\newcommand{\graph}{\text{gr}}

\newcommand{\emptybinding}{\ensuremath{\set{\mu_\emptyset}}}
\newcommand{\emptygraph}{\ensuremath{G_{\emptyset}}}

\newcommand{\expr}{\ensuremath{\xi}}

\newcommand{\propaccess}[2]{\ensuremath{#1.#2}} 
\newcommand{\labelcheck}[2]{\ensuremath{#1\!:\!#2}} 

\newcommand{\length}{\text{length}}

\newcommand{\mypar}[1]{\vspace{3mm}\noindent{\bf #1}.}

\newcommand{\fexp}{\text{\it fexpr}}
\newcommand{\pname}{\text{\it pname}}

\acmDOI{}

\acmISBN{}

\acmYear{2017}
\copyrightyear{2017}

\acmArticle{}
\acmPrice{}

\begin{document}
%\title{G-CORE: A Closed and Composable Graph Query Language with Paths as First Class Citizens}
%\title{G-CORE\\ a Closed Query Language on Property Graphs with Paths}
\title{G-CORE\\ A Core for Future Graph Query Languages}
\subtitle{Designed by the LDBC Graph Query Language Task Force}
\subtitlenote{
This paper is the culmination of 2.5 years of intensive discussion between the LDBC Graph Query Language Task Force and members of industry and academia. 
We thank the following organizations who participated in this effort: 
Capsenta,
HP, 
Huawei, 
IBM, 
Neo4j, 
Oracle, 
SAP 
and
Sparsity.
We also thank the following people for their participation:
Alex Averbuch, 
Hassan Chafi,
Irini Fundulaki, 
Alastair Green, 
Josep Lluis Larriba Pey,
Jan Michels, 
Raquel Pau, 
Arnau Prat, 
Tomer Sagi
and  
Yinglong Xia.
}

\author{Renzo Angles}
\affiliation{%
  \institution{Universidad de Talca}
}
\author{Marcelo Arenas}
\affiliation{%
  \institution{PUC Chile}
}
\author{Pablo Barcel\'o}
\affiliation{%
  \institution{DCC, Universidad de Chile}
}
\author{Peter Boncz}
\affiliation{%
  \institution{CWI, Amsterdam}
}
\author{George Fletcher}
\affiliation{%
  \institution{Technische Universiteit Eindhoven}
}
\author{Claudio Gutierrez}
\affiliation{%
  \institution{DCC, Universidad de Chile}
}
\author{Tobias Lindaaker}
\affiliation{%
  \institution{Neo4j}
}
\author{Marcus Paradies}
\affiliation{%
  \institution{SAP SE}
}
\author{Stefan Plantikow}
\affiliation{%
  \institution{Neo4j}
}
\author{Juan Sequeda}
\affiliation{%
  \institution{Capsenta}
}
\author{Oskar van Rest}
\affiliation{%
  \institution{Oracle}
}
\author{Hannes Voigt}
\affiliation{%
  \institution{Technische Universit\"at Dresden}
}

\renewcommand{\shortauthors}{R. Angles et al.}
\renewcommand{\shorttitle}{G-CORE}

\sloppy

\begin{abstract}
We report on a community effort between industry and academia to shape the future of graph query languages. 
We argue that existing graph database management systems should consider supporting a query language with two key characteristics.
First, it should be composable, meaning, that graphs are the input and the output of queries. 
Second, the graph query language should treat paths as first-class citizens. 
Our result is G-CORE, a powerful graph query language design that fulfills these goals, and strikes a careful balance between path query expressivity and evaluation complexity.
 \end{abstract}

%\begin{CCSXML}
%<ccs2012>
%  <concept_id>10010520.10010553.10010562</concept_id>
%  <concept_desc>Computer systems organization~Embedded systems</concept_desc>
%  <concept_significance>500</concept_significance>
%  <concept_id>10010520.10010575.10010755</concept_id>
%  <concept_desc>Computer systems organization~Redundancy</concept_desc>
%  <concept_significance>300</concept_significance>
%  <concept_id>10010520.10010553.10010554</concept_id>
%  <concept_desc>Computer systems organization~Robotics</concept_desc>
%  <concept_significance>100</concept_significance>
%  <concept_id>10003033.10003083.10003095</concept_id>
%  <concept_desc>Networks~Network reliability</concept_desc>
%  <concept_significance>100</concept_significance>
%</ccs2012>  
%\end{CCSXML}
%
%\ccsdesc[500]{Computer systems organization~Embedded systems}
%\keywords{ACM proceedings, \LaTeX, text tagging}

\maketitle

% flatex input: [motivation.tex]
%!TEX root = main.tex

\section*{Preamble}

G-CORE is a design by the LDBC Graph Query Language Task Force, consisting of members from industry and academia, intending to bring the best of both worlds to graph practitioners.

LDBC is {\bf not} a standards body and rather than proposing a new standard, we hope that the design and features of G-CORE will guide the evolution of both existing and future graph query languages, towards making them more useful, powerful and expressive. 

\section{Introduction}

In the last decade there has been increased interest in graph data management.
In industry, numerous systems that store and query or analyze such data have been developed. 
In academia, manifold functionalities for graph databases have been proposed, studied and experimented with.  

Graphs are the ultimate abstraction for many real world processes and today the computer infrastructure exists to collect, store and handle them as such. 
There are several models for representing graphs.
%Graphs are a flexible data model as it does not not impose a schema,
%such that users can incrementally add nodes and edges. %, labels and
%properties. 
%keeping its flexibility as well their ability to represent datasets of
%any form, be it tabular, tree-shaped or of course graph.
Among the most popular is the {\em property graph data model}, which is a directed graph with labels on both nodes and edges, as well as $\langle$property,value$\rangle$ pairs associated with both. 
It has gained adoption with systems such as 
AgensGraph~\cite{agens},
Amazon Neptune~\cite{neptune},
ArangoDB~\cite{arangodb},
Blazegraph~\cite{blazegraph},
CosmosDB~\cite{cosmos},
DataStax Enterprise Graph~\cite{dsegraph},
HANA Graph~\cite{RPBL13},
JanusGraph~\cite{janus},
Neo4j~\cite{N17}, 
Oracle PGX~\cite{sevenich2016using},
OrientDB~\cite{orientdb},
Sparksee~\cite{sparksee},
Stardog~\cite{stardog},
TigerGraph~\cite{tiger},
Titan~\cite{titan}, 
etc.  
These systems have their own storage models, functionalities, libraries and APIs and many have query languages.
This wide range of systems and functionalities poses important interoperability challenges to the graph database industry.
In order for the graph database industry to cooperate, community efforts such as Apache Tinkerpop, openCypher\cite{openCypher} and the Linked Data Benchmark Council (LDBC) are providing vendor agnostic graph frameworks, query languages and benchmarks.

LDBC was founded by academia and industry in 2012~\cite{ABLFNENMKVT14} in order to establish standard benchmarks for such new graph data management systems. 
LDBC has since developed a number of graph data management benchmarks~\cite{OALCGPPB15,KEKFA17,IHNHPMCCSATXNB16} to contribute to more objective comparison among systems, informing prospective users of some of the strong- and weak-points of the various systems before even doing a Proof-Of-Concept study, while providing system engineers and architects clear targets for performance testing and improvement.
LDBC regularly organizes Technical User Community (TUC) meetings, where not only members report on progress of LDBC task forces but also gather requirements and feedback from data practitioners, who are also present.
There have been over 40 graph use-case presentations by data practitioners in these TUC meetings, who
often are users of the graph data management software of LDBC members, such as IBM, Neo4j, Ontotext, Oracle and SAP.
The topics and contents of these collected TUC presentations show that graph databases are being adopted over a wide range of application fields, as summarized in Figure~\ref{fig:ldbc}.
This further shows that the desired graph query language features are 
graph pattern matching (e.g., identification of communities in social networks),
graph reachability (e.g., fraud detection in financial transactions or insurance), 
weighted path finding (e.g., route optimization in logistics, or bottleneck detection in telecommunications),
graph construction (e.g., data integration in Bioinformatics or specialized publishing domains such as legal)  
and 
graph clustering (e.g., on social networks for customer relationship management). 

\begin{figure}[t]
{\small\renewcommand*{\arraystretch}{0.9}
\begin{tabular}{|l|l|c|l|l|}
\cline{1-2} \cline{4-5}
\multicolumn{2}{|c|}{Application Fields}&\ \ \ &
\multicolumn{2}{|c|}{Used Features}\\
\cline{1-2} \cline{4-5}
healthcare / pharma& 14 &&graph reachability   & 36 \\
publishing         & 10 &&graph construction   & 34 \\
finance / insurance& 6  &&pattern matching     & 32 \\
cultural heritage  & 6  &&shortest path search & 19 \\
e-commerce         & 5  &&graph clustering     & 14 \\\cline{4-5}
social media       & 4  &\multicolumn{3}{c}{ }\\
telecommunications & 4  &\multicolumn{3}{c}{ }\\\cline{1-2}
\end{tabular}}
%\vspace*{-3mm}
\caption{Graph database usage characteristics derived from the use-case presentations in LDBC TUC Meetings 2012-2017 (source:
https://github.com/ldbc/tuc\_presentations).}\label{fig:ldbc}
%\vspace*{-7mm}
\end{figure}

\subsection{Three Main Challenges}

The following issues are observed about existing graph query languages.
These observations are based on the LDBC TUC use-case analysis and feedback from industry practitioners:

\mypar{Composability} 
The ability to plug and play is an essential step in standardization. 
Having the ability to plug outputs and inputs in a query language incentivizes its adoption (modularity, interoperability); 
simplify abstractions, users do not have to think about multiple data models during the query process; 
and increases its productivity, by facilitating reuse and decomposition of queries. 
Current query languages do not provide full composability because they output tables of values, nodes or edges.

\mypar{Path as first-class citizens} 
The notion of Path is fundamental for graph databases, because it introduces an intermediate abstraction level that allows to represents how elements in a graph are related.
The facilities provided by a graph query language to manipulate paths (i.e. describe, search, filter, count, annotate, return, etc.) increase the expressivity of the language.
Particularly, the ability to return paths enables the user to post-process paths within the query language rather that in an ad-hoc manner \cite{dries}. 
%Paths are one of the most characteristic feature when working with graphs.
%The ability of a query language to return paths but also store paths, annotate paths (for example with a cost) and query paths increases the expressivity of the language, thus enabling to post-process a path within the language rather that in an ad-hoc manner outside of the query language.

\mypar{Capture the core of available languages}
Both the desirability of a {\em standard} query language and the difficulty of achieving this, is well-established. 
This is particularly true for graph data languages due to the diversity of models and the rich properties of the graph model. 
This motivates our approach to take the successful functionalities of current languages as a base from where to develop the next generation of languages.

\subsection{Contributions}

Since the lack of a common graph query language kept coming up in LDBC benchmark discussions, it was decided in 2014 to create a task force to work on a common direction for property graph query languages.
The authors are members of this task-force.

This paper presents $\gcore$, a closed query language on Property Graphs. % with paths as first-class citizens, which is a proposal for a graph query language which 
It is a coherent, comprehensive and consistent integration of industry desiderata and the leading functionalities found in industry practices and theoretical research. 

\noindent The paper presents the following contributions: 

\mypar{Path Property Graph model} $\gcore$ treats paths as
first-class citizens. This means that paths are outputs of certain queries. The fact
that the language must be closed implies that paths must be part of the graph
data model. This leads to a principled change of the data model: \emph{it
  extends property graphs with paths}. That is, in a graph, there is also a
(possibly empty) collection of paths; where a path is a concatenation of
existing, adjacent, edges. Further, given that nodes, edges and paths are all
first-class citizens, paths have identity and can also have labels and
$\langle$property,value$\rangle$ pairs associated with them.
%Paths have identity, i.e. a graph may contain multiple, distinct paths with the same structure, labels, and properties (similar to how classic property graphs may contain multiple, distinct nodes with the same labels and properties). 
This extended property graph model, called the Path Property Graph model, is
backwards-compatible with the property graph model.

\mypar{Syntax and Semantics of G-CORE} 
A key contribution is the formal definition of $\gcore$. 
This formal definition prevents any ambiguity about the functionality of the language, thus enabling the development of correct implementations. 
In particular, an open source grammar for $\gcore$ is available\footnote{\url{https://github.com/ldbc/ldbc\_gcore\_parser}}.

\mypar{Complexity results} 
To ensure that the query language is practically usable on large data, the design of $\gcore$ was built on previous complexity results.
Features were carefully restricted in such ways that $\gcore$ is tractable (each query in the language can be evaluated efficiently). 
%polynomial in data complexity.
Thus, $\gcore$ provides the most powerful path query functionalities proposed so far, while carefully avoiding intractable %data 
complexity.  %Our results are discussed in Section XX and Appendix~\ref{theory}.

\mypar{Organization of the paper}
This paper first defines the Extended Property Graph model in Section~\ref{sec:datamodel}. Then it explains $\gcore$ in Section~\ref{sec:gcore-by-example} via a guided tour, using examples on the LDBC Social Network Benchmark dataset~\cite{OALCGPPB15}, which demonstrate its main features. 
We summarize our formal contributions, comprising syntax, semantics and complexity analysis of $\gcore$ in Section \ref{sec:formal}, while the details of these are described in Appendix~\ref{theory}. 
In Section~\ref{sec:extensions}, we show how $\gcore$ can be easily extended to handle tabular data. 
We discuss related work in Section~\ref{sec:comparison}, before concluding in Section~\ref{sec:conclusion}.

\begin{figure}[t]

        %\vspace{-1cm}
    \begin{center}
\includegraphics[width=0.6\columnwidth]{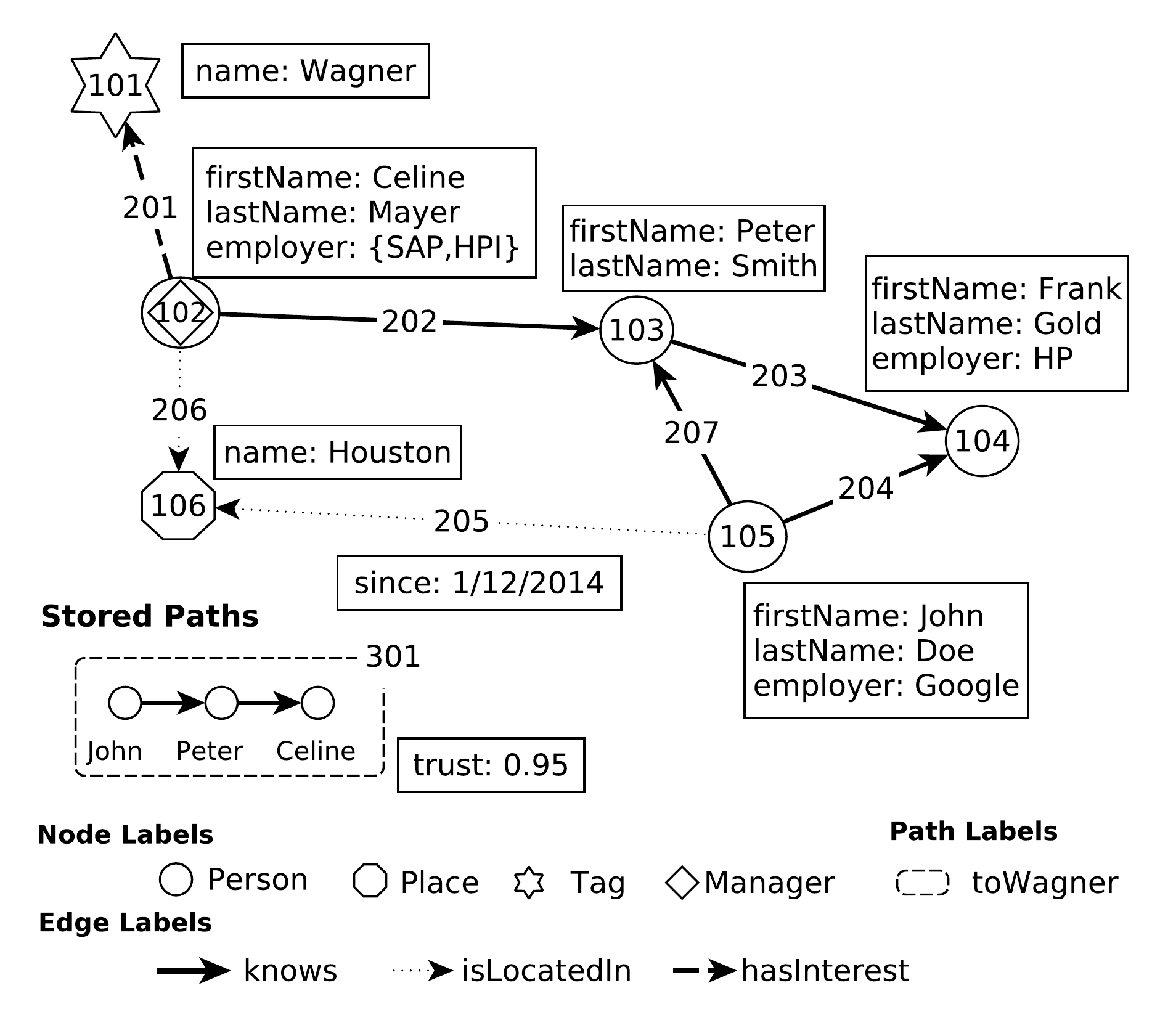}
    \end{center}
        %\vspace{-5mm}
\caption{A small social network. A Path Property Graph (PPG) is a Property Graph that can have ``Stored Paths''.}\label{fig:small-social-network}
%\vspace*{-3mm}
\end{figure}

\section{Path Property Graphs}
\label{sec:datamodel}

We first define the data model of \gcore, which is 
an extension of the Property Graph data model~\cite{RN10,AABHRV16,V17,N17,LPGM17}. 
We call this model the \textit{Path Property Graph} model, or
$\gcgraph$ model for short.
Let $\bL$ be an infinite set of label names for nodes, edges and paths, $\bK$ an infinite set of property names, 
and $\bV$ an infinite set of literals (actual values such as integer and real numbers, strings, dates, 
truth values $\bot$ and $\top$, that represent true and false, respectively,  etc.). 
Moreover, given a set $X$, let $\fset(X)$ denote the set of all finite subsets of $X$ (including the empty set), and $\flist(X)$ denote the 
set of all finite lists of elements from $X$ (including the empty list).

\begin{definition}\label{def-ext-prop-graph}
A {\em $\gcgraph$} is a tuple $G = (N, E, P$, $\rho, \delta, \lambda, \sigma)$, where:
\begin{enumerate}
\item $N$ is a finite set of node identifiers, $E$ is a finite set of edge identifiers and $P$ is a finite set of path identifiers, where $N$, $E$ and $P$ are pairwise disjoint.

\item $\rho : E \to (N \times N)$ is a total function.

\item $\delta : P \to \flist(N \cup E)$ is a total function such that for every $p \in P$, it holds that $\delta(p) = [a_1, e_1, a_2, \ldots, a_n, e_{n}, a_{n+1}]$, where: (i) $n \geq 0$, (ii) $e_j \in E$ for every $j \in \{1, \ldots, n\}$, and (iii) $\rho(e_j) = (a_{j}, a_{j+1})$ or $\rho(e_j) = (a_{j+1}, a_{j})$ for every $j \in \{1, \ldots, n\}$

\item $\lambda : (N \cup E \cup P) \to \fset(\bL)$ is a total function.

\item $\sigma : (N \cup E \cup P) \times \bK \to \fset(\bV)$ is a total function for which there exists a finite set of tuples $(x,k) \in (N \cup E \cup P) \times \bK$ such that $\sigma(x,k) \neq \emptyset$
\end{enumerate}
\end{definition}

Given an edge $e$ in a $\gcgraph$ $G$, if $\rho(e) = (a,b)$, then $a$ is the starting node of $e$ and $b$ is the ending node of $e$. The function $\rho$ allows us to have several edges between the same pairs of nodes. Function $\delta$ assigns to each path identifier $p \in P$ an actual {\em path} in $G$: this is a  list 
$[a_1, e_1, a_2, \ldots, a_n, e_n, a_{n+1}]$ satisfying condition (3) in Definition \ref{def-ext-prop-graph}. 
Function $\lambda$ is used to specify the set of labels of each node, edge, and path, while function $\sigma$ is used to specify the values of a property for every node, edge, and path. To be precise, if $x \in (N \cup E \cup P)$ and $k \in \bK$ is a property name, then $\sigma(x,k)$ is the set of values of the property $k$ for the identifier $x$. Observe that if $\sigma(x,k) = \emptyset$, then we implicitly assume that property $k$ is not defined for identifier $x$, as there is no value of this property for this object. Note that although $\bK$ is an infinite set of property names, in $G$ only a finite number of properties are assigned values as we assume that there exists a finite set of tuples $(x,k) \in (N \cup E \cup P) \times \bK$ such that $\sigma(x,k) \neq \emptyset$. 

\begin{example}
As a simple example of a $\gcgraph$, consider the small social network graph
given in Figure \ref{fig:small-social-network}.  Here we have
\begin{eqnarray*}
    N &=&  \{101, 102, 103, 104, 105, 106\}, \\
    E &=&  \{ 201, 202, 203, 204, 205, 206, 207 \}, \text{and} \\
    P &=&  \{ 301 \}  
\end{eqnarray*}
as node, edge, and path identifiers, respectively; 
\begin{eqnarray*}
    \rho  &=&  \{ 201\mapsto (102, 101),  \ldots, 207\mapsto (105, 103)\} \text{ and}\\
    \delta &=& \{301 \mapsto [105, 207, 103, 202, 102] \} 
\end{eqnarray*}
as edge and path assignments, respectively; and, 
\begin{multline*}
    \lambda = \{101\mapsto \{\text{Tag}\}, 102\mapsto \{\text{Person}, \text{Manager}\}, \ldots, %\\ 
    201\mapsto\{\text{hasInterest}\}, \ldots, 301\mapsto\{\text{toWagner}\}\} 
\end{multline*}%\vspace*{-11mm}
and %\vspace*{8mm}
\begin{multline*}
    \sigma = \{(101, \text{name})\mapsto \set{\text{Wagner}}, \ldots, %\\ 
    (205, \text{since})\mapsto \set{\text{1/12/2014}}, \ldots, (301, \text{trust})\mapsto \set{0.95}\}
\end{multline*}
as label and property value assignments, respectively.
\end{example} 

\mypar{Paths}
It is worth remarking that paths are included as a first-class citizens in this data model (at the level of nodes and edges). 
In particular, paths can have labels and properties, where the latter can be used to describe built-in properties like the length of the path.
In our example above, the path with identifier 301 has label ``toWagner'' and value 0.95 on property ``trust''.

For convenience, we use \nodes{p} and \edges{p} to denote the list of all nodes
and edges of a path bound to a variable $p$, respectively.
Formally, if $\delta(p) = [a_1, e_1, a_2, \ldots, e_n, a_{n+1}]$ then 
$\nodes{p}=[a_1, \ldots, a_{n+1}]$ and $\edges{p}=[e_1, \ldots, e_n]$.
In our example above, $\nodes{301} = [102, 103, 105]$ and
 $\edges{301} = [202, 207]$.

% flatex input end: [data-model.tex]

%\keywords{ACM proceedings, \LaTeX, text tagging}
% flatex input: [gcore-by-example.tex]
%!TEX root = main.tex
\section{A Guided Tour of G-CORE }
\label{sec:gcore-by-example}

\begin{figure}[t]
\centering
\includegraphics[width=0.6\columnwidth]{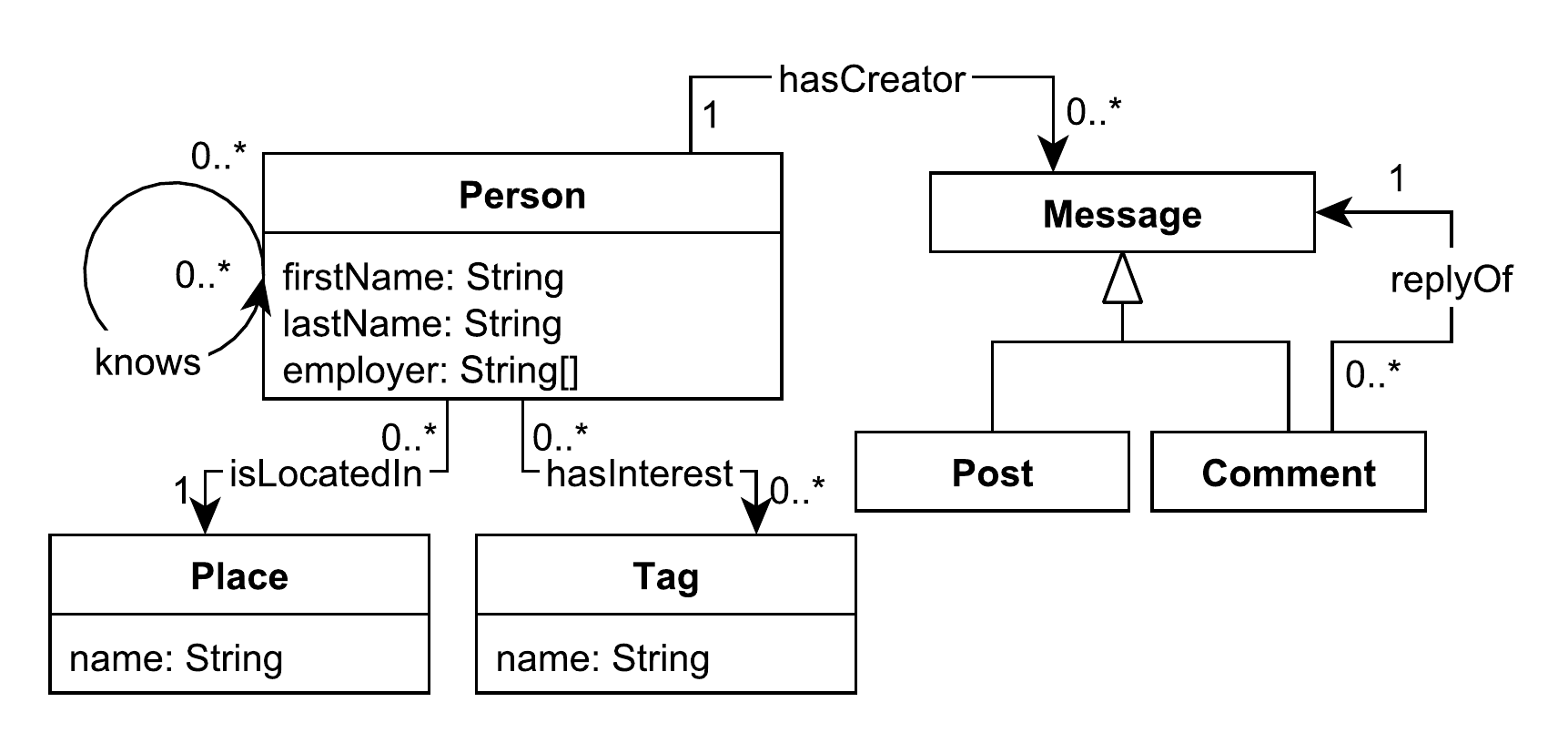}
%\vspace*{-8mm}
\caption{Social Network Benchmark schema (simplified).}\label{fig:snb-schema} %\vspace*{-5mm}
\end{figure}

\begin{figure}[t!]
\includegraphics[width=0.6\columnwidth]{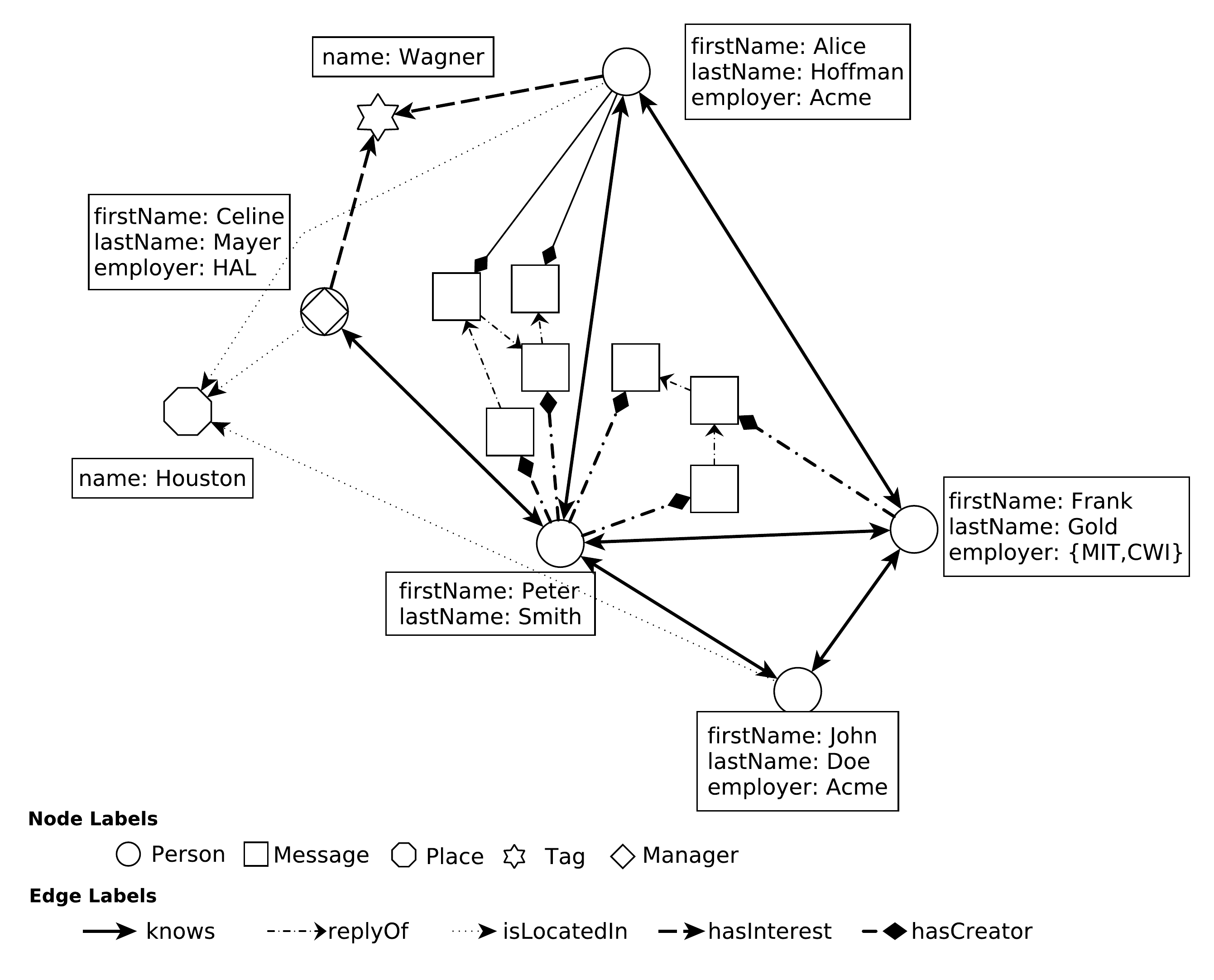}
%\vspace*{-8mm}
\caption{Initial graph database (\lstinline{social_graph}). Note that the \lstinline{knows} edges are drawn bi-directionally -- this means there are two edges: one in each direction.}\label{fig:social-network}%\vspace*{-5mm}
\end{figure}

We will now demonstrate and explain the main features of the $\gcore$ language.
The concrete setting is the LDBC Social Network Benchmark (SNB), as illustrated in the
 simple social network from Figure~\ref{fig:small-social-network}, whose (simplified) schema is 
 depicted in Figure~\ref{fig:snb-schema}.  
Figure~\ref{fig:social-network} depicts the toy instance (which we refer to as \lstinline{social_graph})
 on which our example queries are evaluated.
The use-cases in these examples are data integration and expert finding in a social network. 

\mypar{Always returning a graph} 
Let us start with what is possibly one of the simplest $\gcore$ queries:

\begin{lstlisting}
CONSTRUCT (n)
  MATCH   (n:Person) 
    ON    social_graph
    WHERE n.employer = 'Acme' #\label{line:filtering1}#
\end{lstlisting}

In $\gcore$ every query returns a graph, as embodied by the \lstinline{CONSTRUCT} clause 
 which is at the start of every query body.
This example query constructs a new graph with no edges and only nodes, namely 
 those persons who work at Acme -- all the labels and properties that these person nodes had
 in \lstinline{social_graph} are preserved in the returned result graph.

\mypar{Match and Filter} 
The \lstinline{MATCH..ON..WHERE} clause matches one or more (comma separated) {\em graph patterns} 
 on a named graph, using the homomorphic semantics~\cite{AABHRV16}.

Systems may omit \lstinline{ON} if there is a {\em default} graph -- let us assume in the sequel 
 that \lstinline{social_graph} is the default graph.
Parenthesis demarcate a node, where \lstinline{n} is a variable bound to the identity
 of a node, \lstinline{:Person} a label, and \lstinline{n.employer} a property.
The $\gcore$ builds on the ASCII-art syntax from Cypher~\cite{Cypher18} and the regular path expression syntax from PGQL \cite{RHKMC16}, which
 has proven intuitive, effective and popular among property graph database users.

The previous example contains a \lstinline{WHERE} filter with the obvious semantics: it eliminates 
 all matches where the employer is not~Acme.

\mypar{Multi-Graph Queries and Joins}
A more useful query would be a simple {\em data integration} query, where we might 
 have loaded (unconnected) company nodes into a temporary graph \lstinline{company_graph}, 
 but now want to create a unified graph where employees and companies are connected
 with an edge labeled \lstinline{worksAt}.
Let us assume that \lstinline{company_graph} contains nodes for Acme, HAL, CWI and MIT.
As an aside, the real SNB dataset already contains such Company nodes with \lstinline{:worksAt}
 edges to the employees (which in reality do not have an \lstinline{employer} property).

The below query has a \lstinline{MATCH} clause with two graph patterns, matching these
 on two different input graphs.
Graph patterns that do not have variables in common lead to the Cartesian product of variable 
 bindings, but this query also has a \lstinline{WHERE} clause that turns it into an equi-join:

\noindent\begin{lstlisting}
CONSTRUCT (c)<-[:worksAt]-(n)
  MATCH   (c:Company) ON company_graph, #\label{line:multiplegraph}#
          (n:Person)  ON social_graph
    WHERE c.name = n.employer #\label{line:valuejoin}# #\label{line:filtering2}#
UNION social_graph #\label{line:wholegraph1}#
\end{lstlisting}

The \lstinline{UNION} operator takes it intuitive meaning, and will be touched upon
 later when we talk about node and edge identity.

Generally speaking, \lstinline{MATCH} produces a {\em set of bindings} which
alternatively may be viewed as a table having a column for each variable and one
row for each binding.
Bindings typically contain node, edge and path identities, whose shape is opaque, but 
 we use intuitive names prefixed \# here:

\begin{center}
{\scriptsize\begin{tabular}{|l|l|}
\hline
{\bf c}         & {\bf n}\\
\hline
\hline
{\bf \#Acme} & {\bf \#Alice}\\
{\bf \#HAL}     & {\bf \#Celine}\\
{\bf \#Acme} & {\bf \#John}\\
\hline
\end{tabular}}
\end{center}

\mypar{Dealing with Multi-Valued properties} 
In the previous query there is the complication that \lstinline{n.employer} is {\em multi-valued}
 for Frank Gold: he works for both MIT and CWI.
Therefore, his person node fails to match with both companies.
To explain this, we show the bindings and values of \lstinline{c.name} and \lstinline{n.employer} 
 if \lstinline{WHERE c.name = n.employer} were omitted, and the query would be a Cartesian product: 

\begin{center}
{\scriptsize\begin{tabular}{|l|l|l|l|}
\hline
{\bf c}       & {\bf c.name}  & {\bf n}       & {\bf n.employer}\\
\hline
\hline
\#MIT         & "MIT"         & \#Peter       & \\
\#CWI         & "CWI"         & \#Peter       & \\
\#Acme        & "Acme"        & \#Peter       & \\
\#HAL         & "HAL"         & \#Peter       & \\
\#MIT         & "MIT"         & \#Frank       & \{"CWI",\,"MIT"\} \\
\#CWI         & "CWI"         & \#Frank       & \{"CWI",\,"MIT"\}\\
\#Acme        & "Acme"        & \#Frank       & \{"CWI",\,"MIT"\}\\
\#HAL         & "HAL"         & \#Frank       & \{"CWI",\,"MIT"\}\\
\#MIT         & "MIT"         & \#Alice       & "Acme" \\
\#CWI         & "CWI"         & \#Alice       & "Acme"\\
{\bf \#Acme}  & {\bf "Acme"}  & {\bf \#Alice} & {\bf "Acme"} \\
\#HAL         & "HAL"         & \#Alice       & "Acme"\\
\#MIT         & "MIT"         & \#Celine      & "HAL" \\
\#CWI         & "CWI"         & \#Celine      & "HAL"\\
\#Acme        & "Acme"        & \#Celine      & "HAL"\\
{\bf \#HAL}   & {\bf "HAL"}   & {\bf \#Celine}& {\bf "HAL"}\\
\#MIT         & "MIT"         & \#John        & "Acme" \\
\#CWI         & "CWI"         & \#John        & "Acme"\\
{\bf \#Acme}  & {\bf "Acme"}  & {\bf \#John}  & {\bf "Acme"}\\
\#HAL         & "HAL"         & \#John        & "Acme"\\
\hline 
\end{tabular}}
\end{center}
Notice that according to the definition of our data model, the value of {\bf c.name} is a set. But in the case {\bf c.name}  is a singleton set, we omit curly braces, so we simply write {\small{\tt "MIT"}} instead of {\small{\tt \{"MIT"\}}}.
In the table above, the rows in bold would be the ones that earlier led to bindings surviving the join.
Essentially, \lstinline{"MIT"={"CWI","MIT"}{\small{\tt \}}} and \lstinline{"CWI"={"CWI","MIT"}{\small{\tt \}}} evaluate
to \lstinline{FALSE}.  

Note that Peter is unemployed, so his \lstinline{n.employer} value is unbound.
More precisely, its Person node does not have an employer property at all. 
%In the $\gcore$ data model, properties values cannot be \lstinline{NULL}.
In case of an absent property, its evaluation results in the empty set, 
 which a length test can detect.
$\gcore$ provides \lstinline{CASE} expressions to coalesce such missing data
 into other values.

One way to resolve the failing join for Frank, would be to use \lstinline{IN} instead of 
\lstinline{=}, so the comparisons mentioned earlier resolve to \lstinline{TRUE}:

\noindent\begin{lstlisting}
CONSTRUCT (c)<-[:worksAt]-(n)
  MATCH   (c:Company) ON company_graph,  #\label{line:cartesianproduct}#
          (n:Person)  ON social_graph
    WHERE c.name IN n.employer #\label{line:inlist}#
UNION social_graph #\label{line:wholegraph2}#
\end{lstlisting}

Notice that the \lstinline{IN} operator can be used when \lstinline{c.name} is a singleton set, as in this case it is natural to ask whether the value in \lstinline{c.name} is an element of \lstinline{n.employer}. If we need to compare \lstinline{c.name} with \lstinline{n.employer} as sets, then the operator \lstinline{SUBSET} can be used.

Another way to deal with this in $\gcore$ is to bind a variable (\{\lstinline{name:e}\}) 
 to the \lstinline{employer} property, which unrolls multi-valued properties into individual bindings: 

\noindent\begin{lstlisting}
CONSTRUCT (c)<-[:worksAt]-(n)
  MATCH   (c:Company)             ON company_graph, 
          (n:Person {employer=e}) ON social_graph
    WHERE c.name = e #\label{line:matchliterals1}#
UNION social_graph #\label{line:wholegraph3}#
\end{lstlisting}

Inside the \lstinline{MATCH} expression that binds a node, curly braces can be used to
 bind variables to property values.
The set of bindings for this \lstinline{MATCH} (which includes the join) now has
three variables and the following bindings:

\begin{center}
{\scriptsize\begin{tabular}{|l|l|l|}
\hline
{\bf c}       & {\bf n}       & {\bf e}\\
\hline
\hline
{\bf \#MIT}   & {\bf \#Frank} & {\bf "MIT"}\\
{\bf \#CWI}   & {\bf \#Frank} & {\bf "CWI"}\\
{\bf \#Acme}& {\bf \#Alice} & {\bf "Acme"}\\
{\bf \#HAL}    & {\bf \#Celine} & {\bf "HAL"}\\
{\bf \#Acme}& {\bf \#John}  & {\bf "Acme"}\\
\hline 
\end{tabular}} 
\end{center}

\mypar{Construction that respects identities} 
The \lstinline{CONSTRUCT} operation fills a graph pattern (used as template) for
each binding in the set of bindings produced by the \lstinline{MATCH} clause.
Edges are denoted with square brackets, and can be pointed towards either direction; in this case there is 
no edge variable, but there is an edge label \lstinline{:worksAt}.
Note, to be precise, that \lstinline{CONSTRUCT} by default {\em groups} bindings when creating elements.
Nodes are grouped by node identity, and edges by the combination of source and destination node.
While five new edges are created here, they are between four existing persons and four existing companies
 due to this grouping. 
For instance, the person {\scriptsize\bf \#Frank}, who works for both MIT and CWI, gets two 
 \lstinline{:worksAt} edges, to respectively company {\scriptsize\bf \#MIT} and company 
 {\scriptsize\bf \#CWI}.

In the last line of this example query, we \lstinline{UNION}-ed these new edges with the 
 original graph, resulting in an enriched graph: the original graph plus five edges.
The ``full graph'' query operators like union and difference are defined in terms of node, edge and path identities.
These identities are taken from the input graph(s) of the query.
$\gcore$ is a query language, not an update language. 
Even though \lstinline{CONSTRUCT} allows with \lstinline{SET prop:=val} and \lstinline{REMOVE prop} 
 to change properties and values (a later example will demonstrate \lstinline{SET}), this does not 
 modify the graph database, it just changes the result of that particular query.
The practice of returning a graph that shares (parts of) nodes, edges and paths with its inputs, 
 using this concept of identity, provides opportunities for systems to share memory and storage 
 resources between query inputs and outputs.

A shorthand form for the union operation is to include a graph name directly in the
 comma separated list of \lstinline{CONSTRUCT} patterns, as depicted in the next query:

\noindent\begin{lstlisting}
CONSTRUCT  social_graph, 
    (x GROUP e :Company {name:=e})<-[y:worksAt]-(n) #\label{line:graphaggr}#
  MATCH   (n:Person {employer=e})  #\label{line:matchliterals2}#
\end{lstlisting}

\mypar{Graph Aggregation} 
The above query demonstrates {\em graph aggregation}.
Supposing there would not have been any company nodes in the graph, we 
 might also have created them with this excerpt:

\lstinline{CONSTRUCT (n)-[y:worksAt]->(x:Company}{\small\tt \{}\lstinline{name:=n.employer}{\small\tt \})}

\noindent
However, this unbound destination node \lstinline{x} would create a company
 node for {\em each} binding\footnote{In addition, it would create a company with the 
 {\texttt name} property with the values $\{\texttt{"CWI"}, \texttt{"MIT"}\}$.}.
This is not what we want: we want only one company per unique name.
Graph aggregation therefore allows an explicit \lstinline{GROUP} clause in each graph 
 pattern element.
Thus, in the above query with \lstinline{GROUP e}, we create only one 
 company node for each unique value of \lstinline{e} in the binding set.
Here the curly brace notation is used inside \lstinline{CONSTRUCT} to 
 instantiate the \lstinline{Company.name} property in the newly created nodes.

The set of bindings of our graph aggregation query example has the same variables 
 \lstinline{n} and \lstinline{e} variables of the previous binding set.
The \lstinline{CONSTRUCT} for node expression \lstinline{(n)} groups by node identity
 so instantiates the nodes with identity {\scriptsize\bf \#Frank}, {\scriptsize\bf \#Alice},
{\scriptsize\bf \#Celine} and {\scriptsize\bf \#John} in the query result.
These nodes were already part of \lstinline{social_graph}, so given that the
 \lstinline{CONSTRUCT} is \lstinline{UNION}-ed with that, no extra nodes result.
%\com{Hannes}{regarding \lstinline{(n)}, I would rather not say that new nodes are created, because with may lead the reader to the conclusion that the labels and properties of \lstinline{(n)} only end up in the result graph via \lstinline{UNION social_graph}, which is wrong. Instead, \lstinline{(n)} adds all nodes -- as they are with their labels and properties -- to the result graph that have an identity listed in the \lstinline{n} column of the binding table.}

For the \lstinline{(x GROUP e..)} node expression, \lstinline{CONSTRUCT} groups by \lstinline{e} into
 bindings {\scriptsize\bf "CWI"}, {\scriptsize\bf "MIT"}, {\scriptsize\bf "Acme"}, and
 {\scriptsize\bf "HAL"} and because \lstinline{x} is unbound, it will create four new nodes
 with, say, identities {\scriptsize\bf \#CWI}, {\scriptsize\bf \#MIT}, {\scriptsize\bf \#Acme} and
 {\scriptsize\bf \#HAL}.
For the edges to be constructed, $\gcore$ performs by default grouping of the bindings on the 
 combination of source and destination node, and this results in again five new edges. 

When using bound variables in a \lstinline{CONSTRUCT}, they must be of the 
 right sort: it would be illegal to use \lstinline{n} (a node) in the place of 
 \lstinline{y} (an edge) here. 
In case an {\em edge} variable (here: \lstinline{y}) would have been bound 
 (in the \lstinline{MATCH}), \lstinline{CONSTRUCT} imposes the restriction that its node variables 
 must also be bound, and be bound to exactly its source and destination nodes, because
 changing the source and destination of an edge violates its identity.
However, it can be useful to bind edges in \lstinline{MATCH} and use these to construct edges 
 with a new identity, which are copies of these existing edges in terms of labels and property-values. 
For this purpose, $\gcore$ supports the \lstinline{-[=y]-} syntax which makes a copy of the 
 bound \lstinline{y} edges (as well as the \lstinline{(=n)} syntax for nodes).
Then, the above restriction does not apply.
With the copy syntax, it is even possible to copy all labels and properties of a node to an edge (or a path) and vice versa.
%The \lstinline{CONSTRUCT} grouping then is on the combination of source and destination node bindings (if any) as well as on this bound edge variable. 
%\com{Hannes}{The last sentence is actually not true. The variable in the copy syntax (copy variable) is NOT implicitly part of the group variables. Use case: a may want to copy to multiple edges. However, you still can explicitly mention the copy variable with \lstinline{GROUP} to accomplish the behavior described in the sentence.}

In this example, \lstinline{x} and \lstinline{y} were unbound and could have been omitted.
In the preceding examples, they were in fact omitted.
Unbound variables in a \lstinline{CONSTRUCT} are useful if they occur {\em multiple} times
 in the construct patterns, in order to ensure that the same identities will be used
 (i.e., to connect newly created graph elements, rather than generate independent 
 nodes and edges).

\mypar{Storing Paths with @p} 
$\gcore$ is unique in its treatment of paths, namely as first-class citizens.
The below query demonstrates finding the three shortest paths from John Doe towards each
 other person who lives at his location, reachable over \lstinline{knows} edges, using
 Kleene star notation \lstinline{<:knows*>}:

\noindent\begin{lstlisting}
CONSTRUCT (n)-/@p:localPeople{distance:=c}/->(m) #\label{line:graphproj}#
  MATCH   (n)-/3 SHORTEST p<:knows*> COST c/->(m)#\label{line:topkshortest}#
    WHERE (n:Person) AND (m:Person)
      AND n.firstName = 'John' AND n.lastName = 'Doe' #\label{line:filtering3}#
      AND (n)-[:isLocatedIn]->()<-[:isLocatedIn]-(m) #\label{line:implexist1}#
\end{lstlisting}

In $\gcore$, paths are demarcated with slashes \lstinline{-/ /-}.
In the above example \lstinline{p <:knows*>} binds
 the shortest path between the single node \lstinline{n} (i.e. John Doe)
 and every possible person \lstinline{m}, under the restriction
 that this target person lives in the same place.
By writing e.g., \lstinline{-/3 SHORTEST p <:knows*>/->} we obtain multiple shortest paths 
 (at most 3, in this case) for every source--destination combination; if 
 the number 3 would be omitted, it would default to 1.
In case there are multiple shortest paths with equal cost between two nodes, 
 $\gcore$ delivers just any one of them.
By writing \lstinline{p <:knows*> COST c/->} we bind the shortest path 
 cost to variable \lstinline{c}. 
By default, the cost of a path is its hop-count (length).
We will define weighted shortest paths later.
If we would not be interested in the length, \lstinline{COST c} could be omitted.

\begin{figure}[t]
\centering
\includegraphics[width=0.6\columnwidth]{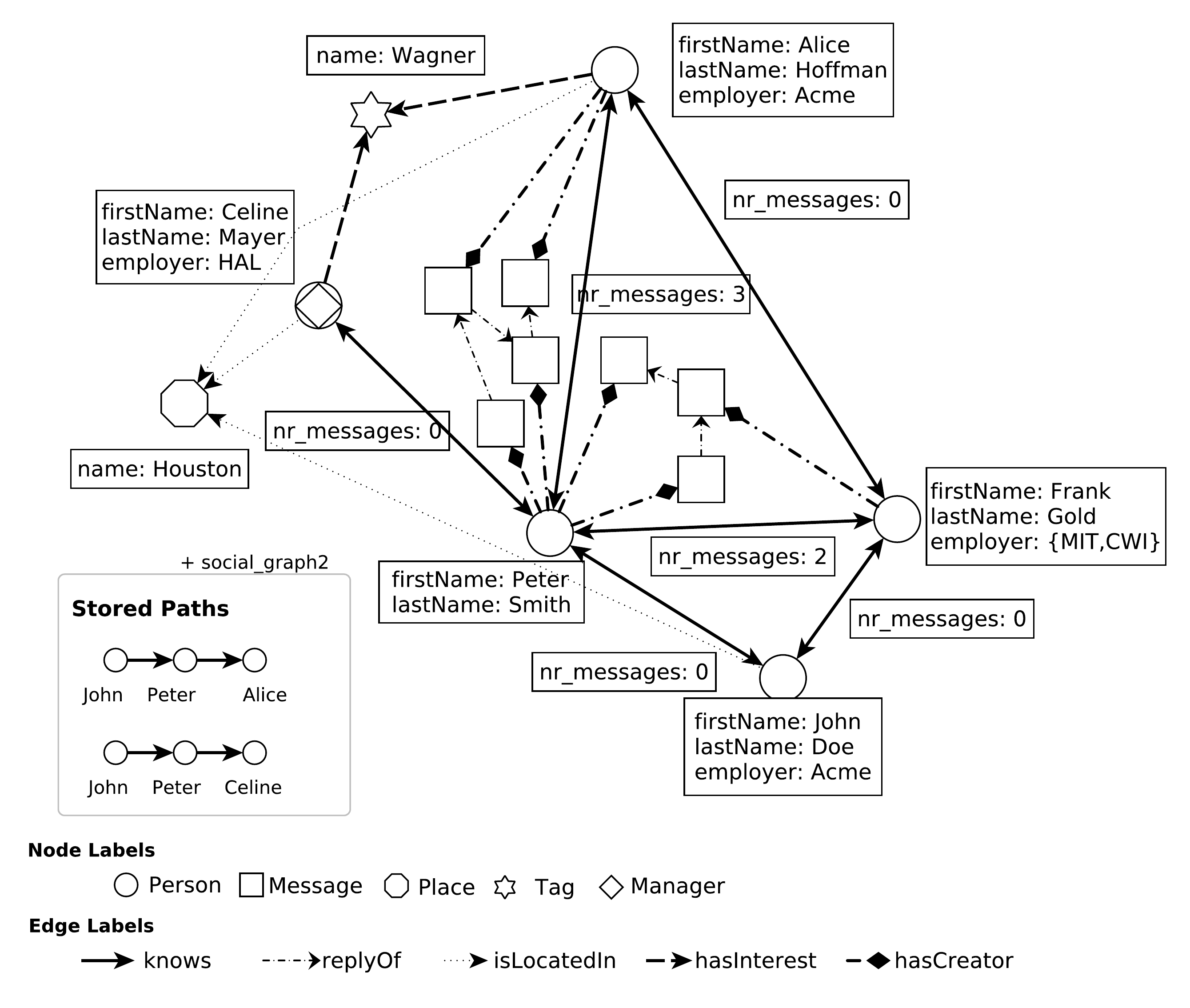}
%\vspace*{-8mm}
\caption{Graph view \lstinline{social_graph1}, which adds \lstinline{nr\_message} properties to the original \lstinline{social_graph} (\lstinline{social\_graph2} is \lstinline{social_graph1} plus the Stored Paths in the grey box).}\label{fig:social-network12}%\vspace*{-5mm}
\end{figure}

In \lstinline{CONSTRUCT (n)-/@p:localPeople}{\small\tt \{}\lstinline{distance:c}{\small\tt \}}, we see the 
 bound path variable \lstinline{@p}.
The \lstinline{@} prefix indicates a {\bf stored path}, that is, this query is 
 delivering a graph with paths.
Each path is stored attaching the label \lstinline{:localPeople},
 and its cost as property \lstinline{distance}.

The graph returned by this query -- which lacks a \lstinline{UNION} with the original 
 \lstinline{social_graph} -- is a {\em projection} of all nodes and edges involved in 
 these stored paths. 
We omitted a figure of this for brevity.

\mypar{Reachability and All Paths} 
In a similar query where we just return \lstinline{m}, and do not store paths,
 the \lstinline{<:knows*>} path expression semantics is a {\em reachability} test:

\noindent\begin{lstlisting}
CONSTRUCT (m)
  MATCH   (n:Person)-/<:knows*>/->(m:Person) #\label{line:allshortest}#
    WHERE n.firstName = 'John' AND n.lastName = 'Doe' #\label{line:filtering4}#
      AND (n)-[:isLocatedIn]->()<-[:isLocatedIn]-(m) #\label{line:implexist2}#
\end{lstlisting}

In this case we use \lstinline{-/<:knows*>/->} without the \lstinline{SHORTEST} keyword. Using \lstinline{ALL} instead of \lstinline{SHORTEST}: asking for {\em all} paths, is not
 allowed if a path variable is bound to it and used somehwere, as this would be intractable or 
 impossible due to an infinite amount of results.
However, $\gcore$ can support it in the case where the path variable is only used to return a graph projection of all paths:

\noindent\begin{lstlisting}
CONSTRUCT (n)-/p/->(m)
  MATCH   (n:Person)-/ALL p<:knows*>/->(m:Person)
    WHERE n.firstName = 'John' AND n.lastName = 'Doe' #\label{line:filtering5}#
      AND (n)-[:isLocatedIn]->()<-[:isLocatedIn]-(m) #\label{line:implexist3}#
\end{lstlisting}

The method~\cite{BLLW12} shows how the materialization of all paths can be avoided by 
 summarizing these paths in a graph projection; hence this functionality is tractable.

\mypar{Existential Subqueries} 
In the SNB graph, \lstinline{isLocatedIn} is not a simple string attribute, 
 but an edge to a city, and the three previous query examples used pattern matching 
 directly in the \lstinline{WHERE} clause:
 \lstinline{(n)-[:isLocatedIn]->()<-[:isLocatedIn]-(m)}. 
$\gcore$ allows this and uses implicit existential quantification, which here is equivalent to:

\noindent\begin{lstlisting}
WHERE EXISTS ( #\label{line:explexist}#
  CONSTRUCT () 
    MATCH (n)-[:isLocatedIn]->()<-[:isLocatedIn]-(m) )
\end{lstlisting}

This constructs one new node (unbound anonymous node variable \lstinline{()}) for
 each match of \lstinline{n} and \lstinline{m} coinciding in the city where
 they are located -- where that city is represented by a \lstinline{()} again. 
Whenever such a subquery evaluates to the empty graph, the automatic existential
 semantics of \lstinline{WHERE} evaluates to \lstinline{FALSE}; otherwise to \lstinline{TRUE}.

%\begin{figure}[t]
%\includegraphics[width=\columnwidth]{figures/social_network1}
%\caption{graph view \lstinline{social_network1}}\label{fig:social-network1}
%\end{figure}

\mypar{Views and Optionals} 
The fact that $\gcore$ is {\em closed} on the $\gcgraph$ data model 
means that subqueries and views are possible. 
In the following example we create such a view:

\noindent\begin{lstlisting}
GRAPH VIEW social_graph1 AS ( #\label{line:graphview1}#
  CONSTRUCT social_graph, 
        (n)-[e]->(m) SET e.nr_messages := COUNT(*) #\label{line:addprop}#
    MATCH (n)-[e:knows]->(m)
      WHERE (n:Person) AND (m:Person)
      OPTIONAL (n)<-[c1]-(msg1:Post|Comment), #\label{line:optionalmatch}#
               (msg1)-[:reply_of]-(msg2),
               (msg2:Post|Comment)-[c2]->(m)
        WHERE (c1:has_creator) AND (c2:has_creator) )
\end{lstlisting}
%\com{Hannes}{The disjunction of labels \lstinline{(x:Post|Comment)} is (a) not explained and (b) break with the principle that everything in the pattern is conjunctive. Every non-conjunctive predicate has to go to \lstinline{WHERE}, here you would have to write \lstinline{(x) WHERE x:Post OR x:Comment}. You are allowed to write \lstinline{(x:Post:Comment)}, but than \lstinline{(x)} needs to have both labels, \lstinline{:Post} and \lstinline{:Comment}. Note that \lstinline{MATCH} and \lstinline{OPTIONAL}, each have their own \lstinline{WHERE}.}

The result of this graph view can be seen in Figure~\ref{fig:social-network12}.
To each \lstinline{:knows} edge, this view adds a \lstinline{nr_messages} property,
 using the \lstinline{SET .. :=} sub-clause.
This sub-clause of \lstinline{CONSTRUCT} can be used to modify properties of nodes, edges and paths that are being constructed.
This particular \lstinline{nr_messages} property contains the amount of messages that 
 the two persons \lstinline{n} and \lstinline{m} have actually exchanged,
 and is a reliable indicator of the intensity of the bond between two persons. 
 % than merely accepting a friend request once.

The edge construction \lstinline{(n)-[e]->(m)} adds nothing new, but as described
 before, performs implicit graph aggregation, where bindings are grouped on \lstinline{n,m,e}, 
 and \lstinline{COUNT(*)} evaluates to the amount of occurrences of each combination.
%\com{Hannes}{In fact, bindings are grouped by \lstinline{n,m,e}, in case you want to be precise. There may (theoretically) be multiple \lstinline{:knows} edges between two persons. In that case, each \lstinline{e} would get the same count. The reason for that is, though, that \lstinline{e} does not show up in the \lstinline{OPTIONAL} pattern, so that the left outer join between \lstinline{MATCH} and \lstinline{OPTIONAL} is only over \lstinline{n,m} and in effect every \lstinline{e} gets joined with every ``message'' between \lstinline{n,m}.}

This example also demonstrates \lstinline{OPTIONAL} matches, such that people who know each
  other but never exchanged a message still get a property \lstinline{e.nr_messages=0}. 
All patterns separated by comma in an \lstinline{OPTIONAL} block must match.
Technically, the set of bindings from the main \lstinline{MATCH} is left outer-joined with 
 the one coming out of the \lstinline{OPTIONAL} block (and there may be more than one 
 \lstinline{OPTIONAL} blocks, in which case this repeats).
There can be multiple \lstinline{OPTIONAL} blocks, and % but they cannot be nested inside each other. Each 
each 
\lstinline{OPTIONAL} block can have its own \lstinline{WHERE}; we demonstrate this here
 by moving some label tests to \lstinline{WHERE} clauses (on \lstinline{:Person} and \lstinline{:has_creator}).
This query also demonstrates the use of disjunctive label tests (\lstinline{msg1:Post|Comment}).

If a query contains multiple \lstinline{OPTIONAL} blocks, they have to be evaluated from the top to the bottom. For example, to evaluate the following pattern:
\noindent\begin{lstlisting}
MATCH (n:Person) 
   OPTIONAL (n)-[:worksAt]->(c) 
   OPTIONAL (n)-[:livesIn]->(a)
\end{lstlisting}
we need to perform the following steps: evaluate \lstinline{(n:Person)} to
generate a binding set $T_1$, 
evaluate \lstinline{(n)-[:worksAt]->(c)} to generate a binding set $T_2$, compute the left-outer join of $T_1$ with $T_2$ to generate a binding set $T_3$,  evaluate \lstinline{(n)-[:livesIn]->(a)} to generate a binding set $T_4$, and compute the left-outer join of $T_3$ with $T_4$ to generate a binding set $T$ that is the result of evaluating the entire pattern. Obviously, in this case the order of evaluation is not relevant, and the previous pattern is equivalent to:
\noindent\begin{lstlisting}
MATCH (n:Person) 
   OPTIONAL (n)-[:livesIn]->(a)
   OPTIONAL (n)-[:worksAt]->(c) 
\end{lstlisting}
However, the order of evaluation can be relevant if the optional blocks of a pattern shared some variables that are not mentioned in the first pattern. For example, in the following expression the variable \lstinline{a} is mentioned in the optional blocks but not in the first pattern \lstinline{(n:Person)}:
\noindent\begin{lstlisting}
MATCH (n:Person) 
   OPTIONAL (n)-[:worksAt]->(a) 
   OPTIONAL (n)-[:livesIn]->(a)
\end{lstlisting}
Arguably, such a pattern is not natural, and it should not be allowed in practice. By imposing the simple syntactic restriction that variables shared by optional blocks have to be present in their enclosing pattern, one can ensure that the semantics of a pattern with multiple \lstinline{OPTIONAL} blocks is independent of the evaluation order~\cite{PAG09}.

%This definition has the advantage that its semantics are independent of the evaluation 
% order of the optional blocks~\cite{something}, which is easier to understand than 
% \lstinline{OPTIONAL} semantics where this is not the case~\cite{somethingelse}.

\mypar{Weighted Shortest Paths} 
The finale of this section describes an example of {\em expert finding}:
 let us suppose that John Doe wants to go to a Wagner Opera, but none of his friends likes Wagner.
He thus wants to know which friend to ask to introduce him to a true Wagner lover who lives in his city
 (or to someone who can recursively introduce him).
To optimize his chances for success, he prefers to try ``friends'' who actually communicate with each other. 
Therefore we look for the {\em weighted} shortest path over the \lstinline{wKnows} (``weighted knows'') 
 {\em path pattern} towards people who like Wagner, where the weight is the inverse of the number of messages 
 exchanged: the more messages exchanged, the lower the cost (though we add one to the divisor to avoid 
 overflow).
For each Wagner lover, we want a shortest path.

In $\gcore$, weighted shortest paths are specified over {\em basic path patterns}, defined
 by a \lstinline{PATH .. WHERE .. COST} clause, because this allows to specify a cost value 
 for each traversed path pattern.
The specified cost must be numerical, and larger than zero (otherwise a run-time error will be raised), 
 where the full cost of a path (to be minimized) is the sum of the costs of all path segments.
If the \lstinline{COST} is omitted, it defaults to 1 (hop count).
%The \lstinline{wKnows} used here is just a single \lstinline{knows} edge, but its definition could also consist of multiple adjacent edges, or even reference other path patterns.  (in which case their \lstinline{COST} is implicitly added to the path pattern cost).
%\com{Hannes}{``implicitly added'', how can that be? That seems not be well defined or at best arbitrarily defined. IMHO it would be better to alias the cost with \lstinline{COST .. AS myCost} so that the cost value shows up as a property of the resulting path object. A basic path pattern \lstinline{a} that uses another path pattern \lstinline{b} can than bind a variable to the path matched by \lstinline{b} with \lstinline{-/p<~b*>/-} and refer to \lstinline{b}'s cost in its own cost function by \lstinline{COST .. p.myCost ..}. This way, that user is very free to define the cost and even more importantly know what is actually going on. If \lstinline{AS ..} is not given for \lstinline{b}, the cost does not show up as a property and cannot contribute to the cost of \lstinline{a}.}

\noindent\begin{lstlisting}
GRAPH VIEW social_graph2 AS ( #\label{line:graphview2}#
PATH wKnows = (x)-[e:knows]->(y)  #\label{line:pathwithfilter}#
  WHERE NOT 'Acme' IN y.employer  #\label{line:filtering6}#
  COST 1 / (1 + e.nr_messages) #\label{line:weightedshortest}#
CONSTRUCT social_graph1, (n)-/@p:toWagner/->(m)
  MATCH   (n:Person)-/p<~wKnows*>/->(m:Person) 
    ON    social_graph1
    WHERE (m)-[:hasInterest]->(:Tag {name='Wagner'}) #\label{line:filtering7}#
      AND (n)-[:isLocatedIn]->()<-[:isLocatedIn]-(m) 
      AND n.firstName = 'John' AND n.lastName = 'Doe')
\end{lstlisting}

%\begin{figure}[ht]
%\includegraphics[width=\columnwidth]{figures/social_network2}
%\caption{graph view \lstinline{social_network2}}\label{fig:social-network2}
%\end{figure}

% flatex input: [features_table.tex]
{\small\begin{table}
\renewcommand*{\arraystretch}{0.9}
\centering
\begin{tabular}{|l|l|}
\hline
\multicolumn{2}{|c|}{\bf Matching}\\
\hline  
Matching all patterns (Homomorphism) & * \\
Matching literal values & \ref{line:matchliterals1}, \ref{line:matchliterals2} \\
Matching $k$ shortest paths & \ref{line:topkshortest} \\
Matching all shortest paths & \ref{line:allshortest} \\
Matching weighted shortest paths & \ref{line:weightedshortest} \\
(multi-segment) optional matching & \ref{line:optionalmatch} \\
\hline
\multicolumn{2}{|c|}{\bf Querying}\\
\hline  
Querying multiple graphs & \ref{line:multiplegraph} \\
Queries on paths & \ref{line:pathquerying} \\
Filtering matches &
\ref{line:filtering1},\ref{line:filtering2},\ref{line:inlist},\ref{line:matchliterals1},\ref{line:filtering3},\ref{line:filtering4},\ref{line:filtering5},\ref{line:filtering6},\ref{line:filtering7},\ref{line:filtering8}
\\
Filtering path expressions & \ref{line:pathwithfilter} \\
Value joins & \ref{line:valuejoin} \\
Cartesian product & \ref{line:cartesianproduct} \\
List membership & \ref{line:inlist} \\
\hline
\multicolumn{2}{|c|}{\bf Subqueries}\\
\hline
Set operations on graphs & \ref{line:valuejoin}, \ref{line:wholegraph2}, \ref{line:wholegraph3} \\
Existential subqueries & \\
- Implicit & \ref{line:implexist1}, \ref{line:implexist2}, \ref{line:implexist3} \\
- Explicit & \ref{line:explexist} \\
\hline
\multicolumn{2}{|c|}{\bf Construction}\\
\hline
Graph construction & * \\
Graph aggregation & \ref{line:graphaggr} \\
Graph projection & \ref{line:graphproj} \\
Graph views & \ref{line:graphview1}, \ref{line:graphview2} \\
Property addition & \ref{line:addprop} \\
\hline
\end{tabular}
\caption{Overview of \gcore{} features and their line occurrences in the example queries in Section~\ref{sec:gcore-by-example}.}
%\vspace*{-3mm}
\label{tab:features}
%\vspace*{-4mm}
\end{table}}

% flatex input end: [features_table.tex]

%\end{figure}

The result of this graph view (\lstinline{social_graph2}) was already depicted in 
 Figure~\ref{fig:social-network12}: it adds to \lstinline{social_graph1} two stored paths.
Apart from \lstinline{GRAPH VIEW name AS (query)}, and similar to CREATE VIEW in SQL  which introduces
 a global name for a query expression, $\gcore$ also supports a \lstinline{GRAPH name AS (query1)} \lstinline{query2} 
 clause which, similar to WITH in SQL, introduces a name that is only visible inside \lstinline{query2}.

\mypar{Powerful Path Patterns} 
Basic \lstinline{PATH} patterns are a powerful building block that allow complex path expressions
 as concatenations of these patterns~\cite{RHKMC16} using a Kleene star, yet still allow for fast 
 Dijkstra-based evaluation.
In $\gcore$, these path patterns can even be non-linear shapes, as \lstinline{PATH}
 can take a comma-separated list of multiple graph patterns. 
But, the path pattern must contain a start and end node (a {\em path segment}), which is taken to be 
 the first and last node in its first graph pattern.
This ensures path patterns can be stitched together to form paths -- a path pattern always contains a path segment between its start and end nodes.
These basic path patterns can also contain \lstinline{WHERE} conditions, without restrictions
 on their complexity.
As John Doe wants his preference for Wagner to remain unknown at his work, we exclude employees of 
 Acme from occurring on the path\footnote{Note that non-linear path patterns, such as \lstinline{PATH (a)-[]-(b),(b)-(c)} add power over linear patterns with existential filters: \lstinline{PATH (a)-[]-(b) WHERE (b)-(c)}, because the latter cannot bind variables. In $\gcore$, variable \lstinline{c} can be used in a \lstinline{COST} expression.}.
The result of this query is a view \lstinline{social_graph2} in which all these shortest paths from
 John Doe to Wagner lovers have been materialized (the \lstinline{@p} in \lstinline{(n)-/@p:toWagner/->(m)}).

A unique capability of $\gcore$ is to query and analyze databases of potentially many stored paths.
We demonstrate this in the final query, where we score John's friends for their aptitude:

\noindent\begin{lstlisting}
CONSTRUCT (n)-[e:wagnerFriend {score:=COUNT(*)}]->(m) 
          WHEN e.score > 0
  MATCH   (n:Person)-/@p:toWagner/->(), (m:Person) #\label{line:pathquerying}#
    ON    social_graph2
    WHERE n = nodes(p)[1] #\label{line:filtering8}#
\end{lstlisting}
%\com{Hannes}{It is complete unclear what \lstinline{my_path} is and where is comes from.}
For the \lstinline{:toWagner} paths, we use \lstinline{nodes(p)[1]} to look at the second node in each path, i.e. a direct friend of John Doe.
$\gcore$ starts counting at 0 so \lstinline{nodes(my_path)[n]} returns the $n-1$ item from the list returned by the function \lstinline{nodes()}, which returns all nodes on a path.
For these direct friends we count how often they occur as the start of \lstinline{:toWagner} paths.
These scores has been attached as a \lstinline{score} property to new \lstinline{:wagnerFriend} edges.
Since in the toy example there are only two Wagner lovers and thus two shortest paths to them,
 both via Peter, the result of this query is a single \lstinline{:wagnerFriend} edge 
 between John and Peter with score 2.

%Table~\ref{tab:features} summarizes the main features of \gcore{} and points to their usage in the example queries (referring to the line numbers).

% flatex input end: [gcore-by-example.tex]

%\keywords{ACM proceedings, \LaTeX, text tagging}
% flatex input: [formalization.tex]
%!TEX root = main.tex

\section{Formalizing and Analyzing G-CORE}
\label{sec:formal} 
%\allowdisplaybreaks

One of the main goals of this paper is to provide a formal definition of the syntax and semantics of a graph query language including the features shown in the previous sections. Formally, a $\gcore$ query is defined by the following top-down grammar:
%\vspace*{-2mm}
\begin{align*}
\query   ::=&\ \hclause \ \fgquery\\
\hclause ::=&\ \varepsilon \ \mid \ \pclause \ \hclause \ \mid \ \gclause \ \hclause \\ 
\fgquery ::=&\ \bgquery \ \mid \\
            &\ (\fgquery \ \sopt  \ \fgquery)\\
\sopt    ::=&\ \Union \ \mid \ \Intersect \ \mid \ \Minus \\ 
\bgquery ::=&\ \cclause \ \mclause
\end{align*}
Thus, a $\gcore$ query consists of a sequence of \Path{} and \Graph{} clauses, followed by a full graph query, i.e., a combination of basic graph queries 
under the set 
operations of union, intersection and difference. 
%The GRAPH clause introduces a synonym for the result of a full graph query ($\fgquery$) that is respectively 
%visible as a named graph $\gid$ in all subsequent queries (\lstinline{GRAPH VIEW}), or only in the 
%directly following query $\query$ (\lstinline{GRAPH}), i.e. $\gclause ::=$
%\vspace*{-2mm}
%\begin{eqnarray*}
%\gclause & ::=  & \kw{Graph}\ \gid\ \kw{As}\ (\fgquery)\ \query\\
%\gv & ::=  & \kw{Graph View}\ \gid\ \kw{As}\ (\fgquery)
%\end{eqnarray*}
A basic graph query consists of a single \construct{} clause followed by one \match{} clause. We have seen 
examples of all these features in Section \ref{sec:gcore-by-example}. 

%<<<<<<< HEAD
%In Appendix \ref{theory}, we provide formal definitions of the syntax and semantics of $\gcore$. As mentioned in Section \ref{sec:gcore-by-example}, the building block in the definition of the semantics is the notion of binding, which is a function that assigns to a each variable in a query either a node or an edge or a path or an actual value. A set of bindings is called a biding table, which is the result of evaluating the graph patterns in a \match{} clause, and which can be filtered by using the conditions in a \where{} clauses, In Appendix \ref{theory}, the notion of binding and binding table are formalized, and the evaluation of a \match{} clause $\varphi$ over a $\gcgraph$ $G$, denoted by $\eval{\varphi}{G}$, is also formalized. Noticed that $\eval{\varphi}{G} = \Omega$, where $\Omega$ is a binding table. In Appendix \ref{theory}, we also formalized the semantics of the \construct{} clause, which given a binding table produces a $\gcgraph$. That is, the evaluation of a \construct{} clause $\psi$ over a binding table $\Omega$ and a $\gcgraph$ $G$, denoted by $\eval{\psi}{\Omega,G}$, is formalized in Appendix \ref{theory}, where $\eval{\psi}{\Omega,G}$ is a $\gcgraph$ $G'$. It is important to notice that $G$ is also an input in the evaluation of $\psi$ as the binding table $\Omega$ can make reference to objects whose labels and properties are stored in $G$. Finally, the other components of the grammar, such as \path{} and \match{} clauses and the operators union, intersection and difference, are also formally defined in Appendix \ref{theory}.
%=======
In Appendix \ref{theory}, we provide formal definitions of the syntax and semantics of $\gcore$. 
%As mentioned in Section \ref{sec:gcore-by-example}, the building block in the definition of the semantics is 
%the notion of {\em binding}, which is a function that assigns to a each variable in a query either a node or an edge or a 
%path or an actual value. 
The basic idea of the language is, given a $\gcgraph$ $G$, to create a new $\gcgraph$ $H$ using the \construct{} clause. This is achieved, in turn, by applying 
an intermediate step provided by the \match{} clause. The application of such a clause 
creates a set of bindings $\Omega$, based on a graph pattern that is evaluated over $G$. 
The interaction between the \match{} and the \construct{} clause  
is explained in more detail below: 

\begin{itemize} 
%semantics of the language centers around the interaction of 
\item 
The result of evaluating the graph pattern $\varphi$ that defines the content of  
a \match{} clause over a $\gcgraph$ $G$ always corresponds to a set $\Omega$ 
of bindings, which is denoted by $\eval{\varphi}{G}$.  
The bindings in $\Omega$ can then be filtered by using boolean conditions specified in the 
\where{} clause. 
\item 
%In Appendix \ref{theory}, the notion of binding and binding table are formalized, and the evaluation of a 
%\match{} clause $\varphi$ over a $\gcgraph$ $G$, denoted by $\eval{\varphi}{G}$, is also formalized. 
%Noticed that $\eval{\varphi}{G} = \Omega$, where $\Omega$ is a binding table. 
A \construct{} clause $\psi$ then takes as input both the $\gcgraph$ $G$ and the
set of bindings
$\Omega$, and produces a new $\gcgraph$ $H$, which is 
denoted by $\eval{\psi}{\Omega,G}$. 
%That is, the evaluation of a \construct{} clause $\psi$ over a binding table $\Omega$ and a $\gcgraph$ $G$, 
%denoted by $\eval{\psi}{\Omega,G}$, is formalized in Appendix \ref{theory}, where $\eval{\psi}{\Omega,G}$ is a $\gcgraph$
%$G'$. 
Note that $G$ is also an input in the evaluation of $\psi$, as the set of bindings $\Omega$ can make reference 
to objects whose labels and properties are defined in $G$. 
\end{itemize} 

The role of the \Path{} clause is to define complex path expressions, as well as the cost associated with them, 
that can in turn be used in graph patterns in the \match{} clause. In this way, it is  
possible to define rich navigational patterns on graphs
that capture expressive query languages that have been studied in depth in the theoretical community (e.g., the class of {\em regular queries} \cite{RRV17}).  

\mypar{Complexity analysis} The G-CORE query language has been carefully designed to ensure that G-CORE queries can be evaluated efficiently 
in {\em data complexity}. Formally, this means that for each fixed G-CORE query $q$, the result $\eval{q}{G}$ of evaluating $q$ over an input 
$\gcgraph$ $G$ can be computed in polynomial time. The main reasons that explain this fact are given below. 

First of all, graph patterns correspond (essentially) to conjunctions of atoms expressing that two nodes are linked by a path satisfying a certain regular expression 
over the alphabet of node and edge labels. The set $\Omega$ of all bindings of a fixed graph pattern $\varphi$ over the input $\gcgraph$ $G$ 
can then be easily computed in polynomial time: we simply look for all possible ways of replacing node and edge 
variables in $\varphi$ by node and edge identifiers in $G$, respectively, and then for 
each path variable $\pi$ representing a path in $G$ from node $u$ to node $v$ whose label 
must conform to a regular expression $r$, we replace $\pi$ by the shortest/cheapest 
path in $G$ from $u$ to $v$ that satisfies $r$ (if it exists). This can be done in polynomial 
time by applying standard automata-theoretic techniques in conjunction with Dijkstra-style algorithms. (Notice that the latter would not be true
if our semantics was based on simple paths; in fact, checking if there is a simple path 
in an extended property graph whose label satisfies a fixed regular expression is an NP-complete problem \cite{MW95}).

Suppose, now, that we are given a fixed G-CORE query $q$ that 
corresponds to a sequence of \path{} clauses followed by a full graph query $q'$. Each \path{} clause is defined by a graph 
pattern $\varphi$ whose evaluation corresponds to a binary relation over  
the nodes of the input $\gcgraph$ $G$. By construction,  
the graph pattern $\varphi$ might mention binary patterns which are defined in previous \path{} clauses. 
Therefore, it is possible to iteratively evaluate 
in polynomial time all graph patterns $\varphi_1,\dots,\varphi_k$ that are mentioned in the 
\path{} clauses of $q$. Once this process is finished, 
we proceed to evaluate $q'$ (which is defined in terms of the $\varphi_i$'s). 

By definition, $q'$ is a boolean combination of full graph queries $q_1,\dots,q_m$. It is thus sufficient to explain how to evaluate 
each such a full graph query $q_j$ in polynomial time. We can assume by construction that $q_j$ consists of a \construct{} clause applied
over a \match{} clause. We first explain how the set of bindings that satisfy the \match{} clause can be computed in polynomial time. 
Since one or more \optional{} clauses could be applied over the \match{} clause, the semantics is based  
on the set $\Omega$ of {\em maximal} bindings for the whole expression, i.e., those that 
satisfy the primary graph pattern expressed in the \match{} clause, and as many atoms as possible 
from the basic graph  patterns that define the \optional{} clauses. The computation of 
$\Omega$ can be carried out in polynomial time by a straightforward extension of the aforementioned 
techniques for efficiently evaluating basic graph patterns. 
Finally, filtering $\Omega$ in accordance with the boolean conditions expressed 
in the \where{} clause can easily be done in polynomial time 
(under the reasonable assumption that such conditions can be evaluated efficiently). Recall that a possible 
 such a condition is \Exists{} $Q$, for $Q$ a subquery. We then need to check whether the evaluation of $Q$ over $G$ yields 
 an empty graph. We inductively assume the existence of an efficient algorithm for checking this. 

% of one or more \optional{} clauses 
Finally, the application of the \construct{} clause on top $G$ and the set $\Omega$ of bindings generated by the \match{} clause 
can be carried out in polynomial time. Intuitively, 
this is because the operations allowed in the \construct{} clause are defined 
by applying some simple aggregation and grouping functions on top of bindings generated by relational algebra operations. 

Given that all evaluation steps of $\gcore$ have polynomial complexity in data size, we conclude that $\gcore$ is tractable. 

%and \match{} clauses 
%and the operators union, intersection and difference, are also formally defined in Appendix \ref{theory}.
%>>>>>>> optional

% flatex input end: [formalization.tex]

%\keywords{ACM proceedings, \LaTeX, text tagging}
% flatex input: [extensions.tex]
%!TEX root = main.tex

\section{Extensions of $\gcore$}
\label{sec:extensions}

%\gcore{} is intentionally desiged as a small language that provides a kernel of graph matching and construction funcionality. 
Practical use of graphs often requires handling tabular data. This suggests that extending \gcore{} with
additional functionality for projecting tabular results or constructing graphs
from imported tabular data may be useful.

\mypar{Projecting tabular results}
In order to integrate the results of graph matching into another system, it
would be necessary or at least convenient to be able to produce a tabular
projection from a query. It is quite straightforward to imagine the
set of bindings produced by \lstinline{MATCH} as a table, and use that
to return a tabular projection.
To achieve this, \gcore{} could be extended with a \lstinline{SELECT} clause for
projecting expressions into a table. Such a tabular projection clause would also
allow the introduction of other common relational operations for slicing,
sorting, and aggregation, similar to Cypher's \verb|RETURN| clause or the
\verb|SELECT| clauses of SQL or SPARQL. Furthermore, \lstinline{SELECT} could be
used for adding various forms of expression-level subqueries, such as scalar
subqueries or list subqueries.

Consider this example of a query that uses tabular projection:

\noindent\begin{lstlisting}
SELECT m.lastName + ', ' + m.firstName AS friendName
MATCH (n:Person)-/<:knows*>/->(m:Person)
    WHERE n.firstName = 'John' AND n.lastName = 'Doe'
      AND (n)-[:isLocatedIn]->()<-[:isLocatedIn]-(m) 
\end{lstlisting}

This query matches persons with the name ``John Doe'' together with
indirect friends that live in the same city and returns a table with the names
of these friends.

It should be noted that the introduction of tabular projection into \gcore{}
changes the language to a multi-sorted language that is capable of either
producing a table or a graph. Such a language would no longer be fully closed
under graphs in a strict sense, which is one reason why this extension has been
left to the future.

\mypar{Importing tabular data}
Conversely, integration of \gcore{} with existing systems raises the question of
how pre-existing tabular data could be processed in a pure graph query language.
Next we present two different alternative proposals for how tabular data could
be brought in to the graph world of \gcore{}.

\mypar{Binding table inputs} One way to import tabular data would be through
the introduction of a new \lstinline{FROM <table>} clause that would import
sets of scalar bindings from a table, which could be used for defining
a graph using the \lstinline{CONSTRUCT} clause such as in this example:

\noindent\begin{lstlisting}
CONSTRUCT
  (cust GROUP custName :Customer {name:=custName}),
  (prod GROUP prodCode :Product {code:=prodCode}),
  (cust)-[:bought]->(prod)
FROM orders
\end{lstlisting}

This will construct a new graph from an input table of customer names
\emph{custName} and product codes \emph{prodCode} by connecting per-customer and
per-product nodes as given by the table.

\mypar{Interpreting tables as graphs} Another alternative is to allow the
\lstinline{MATCH .. ON ..} to treat a tabular input following \lstinline{ON}
as a graph consisting of only isolated nodes that correspond to each row in the
table. The properties of these nodes are the columns of the table and the
values are the fields of the corresponding row.

If we express the previous example using this syntax, it would now look as
follows:

\noindent\begin{lstlisting}
CONSTRUCT
  (cust GROUP o.custName :Customer {name:=o.custName}),
  (prod GROUP o.prodCode :Product {code:=o.prodCode}),
  (cust)-[:bought]->(prod)
MATCH (o) ON orders
\end{lstlisting}

%Since the interaction with the outside world is not yet fully explored, 
%These alternatives are only provided as an illustration of showing that it would be
%possible to extend $\gcore$. We realise that there are other alternatives worth considering.

% flatex input end: [extensions.tex]

%\keywords{ACM proceedings, \LaTeX, text tagging}
% flatex input: [comparison.tex]
%!TEX root = main.tex

\section{Discussion and Related Work}
\label{sec:comparison}
%\section{A Comparison with current graph query languages}

%% historical
Graph query languages have been extensively researched in the past decades, and comprehensive surveys are available.
 Angles and Gutierrez~\cite{90004} surveyed GQLs proposed during the eighties and nineties, before the emergence of current (practical) graph database systems.
  Wood~\cite{90524} studied GQLs focusing on their expressive power and computational complexity.
 Angles~\cite{90525} compares graph database systems in terms of their support for essential graph queries.  
  Barcel\'o~\cite{90852} studies the expressiveness and complexity of several navigational query languages.
 Recently, Angles et al.~\cite{AABHRV16} presented a study on fundamental graph querying functionalities (mainly graph patterns and navigational queries) and their implementation in modern graph query languages.

  The extensive research on querying graph databases has not give rise yet
to a standard query language for property graphs (like SQL for the
relational model). Nevertheless, there are several industrial graph database products on the market.   
 %Blueprints~\cite{91154} was among the first libraries created for the property graph data model. Blueprints is analogous to the JDBC, but for graph databases. 
 Gremlin~\cite{91037} is a graph-based programming language for property graphs which makes extensive use of XPath to support complex graph traversals.  
 Cypher~\cite{91040}, originally introduced by Neo4j and now implemented by a
number of vendors, is a declarative query language for property
graphs that has graph patterns and path queries as basic constructs.
% For our comparisons with Cypher, 
We primarily consider version 9 of Cypher as outlined by
\cite{Cypher18, Cypher9}, while recognizing that Cypher is an evolving
language where several advancements compared to Cypher~9 have already been made.
 Oracle has developed PGQL~\cite{RHKMC16}, a graph query languages that is
closely aligned to SQL and that supports powerful regular path expressions.
Several implementations of PGQL, both for non-distributed~\cite{sevenich2016using} and distributed systems~\cite{roth2017pgx}, exist.
Here, we consider PGQL 1.1~\cite{pgql11}, which is the most recent version that is commercially available~\cite{obdsg}.

$\gcore$ has been designed to support most of the main and relevant features provided by Cypher, PGQL, and Gremlin.
Next we describe the main differences among $\gcore$, Cypher, PGQL, and Gremlin based on the query features described in Section \ref{sec:gcore-by-example}.
Some features (e.g. aggregate operators) will not be discussed here as there are not substantial differences from one language to other.

\mypar{Graph pattern queries} 
The notion of basic graph pattern,
i.e. the conjunction of node-edge-node patterns with filter conditions over them,
%  as the basis of almost every graph query language,
  is intrinsically supported by Cypher, PGQL and $\gcore$.
Some differences arise regarding the support for complex graph patterns (i.e. union, difference, optional). 
Both Cypher and $\gcore$ define the UNION operator to merge the results of two graph patterns.
The absence of graph patterns (negation) is mainly supported via existential subqueries. It is expressed in $\gcore$, Cypher and PGQL with the WHERE NOT (EXISTS) clause.
Optional graph patterns can be explicitly declared in $\gcore$ and Cypher with the OPTIONAL clause. PGQL does not support optional graph patterns, although they can be roughly simulated with length-restricted path expressions (see below).
 Although Gremlin is focused on navigational queries, it supports complex graph patterns (including branches and cycles) as the combination of traversal patterns.

\mypar{Path queries}
$\gcore$, Cypher and PGQL support path queries in terms of regular path expressions (i.e. edges can be labeled with regular expressions).
The main difference between Cypher~9 and PGQL is that the closure operator is restricted to a single repeated label / value.
Both Cypher and PGQL support path length restrictions, a feature that although can be simulated using regular expressions,  improves the succinctness of the language.
 Gremlin supports arbitrary or fixed iteration of any graph traversal (i.e. it is more expressive than regular path queries). 
 Similar to Cypher, Gremlin allows specifying the number of times a traversal should be performed.

\mypar{Query output}
The general approach followed by Cypher~9 and PGQL is to return tables with atomic values (e.g. property values).
This approach can be extended such that a result table can contain complex values.
The extension in Cypher~9 allows returning nodes, edges, and paths.
Recent implementations of Cypher have the ability to return graphs
alongside this table~\cite{CAPS, CypherMultiGraph}.
Gremlin also supports returning complete paths as results.
In contrast, $\gcore$ has been designed to return graphs with paths as first class citizens.

\mypar{Query composition}
With the output of a query in $\gcore$ being a graph, it follows naturally that
queries can be composed by querying the output of one query by means of another
query. Neither Cypher~9, PGQL or SPARQL supports this capability. Gremlin
supports creating graphs and then populate them before querying the new graph. A
notable parallel to $\gcore$ is the evolution of Cypher~10, where queries are
composed through the means of ``table-graphs''. Cypher~10 expresses queries
with multiple graphs and a driving table as input, and produces a set of graphs
along with a table as output. This allows Cypher~10 queries to compose both
linearly and through correlated subqueries \cite{Cypher18}.

\mypar{Evaluation semantics}
There are several variations among the languages regarding the semantics for evaluating graph and path expressions. 
In the context of graph pattern matching semantics,
$\gcore$, PGQL, and Gremlin follow the homomorphism-based semantics (i.e. no restrictions are imposed during matching), and Cypher~9 follows a no-repeated-edge
semantics (i.e. two variables cannot be bound to the same term in a given match) to prevent matching of potentially infinite result sets when enumerating all paths of a pattern.
 With respect to the evaluation of path expressions, $\gcore$ uses shortest-path semantics (i.e. paths of minimal length are returned), Cypher~9 implements
 no-repeated-edge semantics (i.e. each edge occurs at most once in the path), and Gremlin follows arbitrary path semantics (i.e. all paths are considered).
 Additionally, Cypher~9 and PGQL allow changing the default semantics by using built-in functions (e.g. \texttt{allShortestPaths}).

\mypar{Expressive power versus efficiency}
A balance between expressiveness and efficiency (complexity of evaluation) means a balance between practice and theory. 
Currently no industrial graph query language has a theoretical analysis of its complexity and, conversely, theoretical results have not been systematically translated into a design.
One of the main virtues of $\gcore$ is that its design is the integration of both sources of knowledge and experience.

\mypar{SPARQL and RDF}
In this paper we concentrated on property graphs, but there are other data models and query languages available. 
A well-known alternative is the Resource Description Framework (RDF), a W3C recommendation that defines a graph-based data model for describing and publishing Web metadata.
 RDF has a standard query language, SPARQL~\cite{sparql10}, which was designed to support several types of complex graph patterns (including union and optional). 
 Its latest version, SPARQL 1.1~\cite{sparql11}, adds support for negation, regular path queries (called \emph{property paths}), subqueries and aggregate operators.
 The path queries support reachability tests, but paths cannot be returned, nor can the cost of paths be computed.
 The evaluation of SPARQL graph patterns follows a homomorphism-based bag semantics, whereas property paths are evaluated using an arbitrary paths semantics~\cite{AABHRV16}.
 SPARQL allows queries that return RDF graphs, however creating graphs consisting of multiple types of nodes (e.g., belonging to different RDF schema classes; having different properties) in one query is not possible as SPARQL lacks flexible graph aggregation: its CONSTRUCT directly instantiates a single binding table without reshaping.
 Such constructed RDF graphs can not be reused as subqueries, that is, for composing queries; nor does the language offer ``full graph'' operations to union or diff at the graph level.
 We think the ideas outlined in $\gcore$ could also inspire further development of SPARQL.

% flatex input end: [comparison.tex]

%\keywords{ACM proceedings, \LaTeX, text tagging}
% flatex input: [conclusion.tex]
%!TEX root = main.tex

\section{Conclusions}
\label{sec:conclusion}

Graph databases have come of age.
The number of systems, databases and query languages for graphs, both commercial and open source, indicates that these technologies are gaining wide acceptance~\cite{agens,neptune,arangodb,blazegraph,cosmos,dsegraph,RPBL13,janus,N17,sevenich2016using,orientdb,sparksee,stardog,tiger,titan}.

At this stage, it is relevant to begin making efforts towards interoperability of these systems. 
A language like $\gcore$ could work as a base for integrating the manifold models and approaches towards querying graphs. 

We defend here two principles we think should be at the foundations of the future graph query languages:
{\em composability}, that is, having graphs and their mental model as departure and ending point and
treat the most popular feature of graphs, namely {\em paths, as first class citizens}. 

The language we present, $\gcore$, which builds on the experiences with working systems, as well as theoretical results, show these desiderata are not only possible, but computationally feasible and approachable for graph users. 
This paper is a call to action for the stakeholders driving the graph database industry. 

%Adopt a semantics based on shortest path in order to have a tractable language

% flatex input end: [conclusion.tex]

%\keywords{ACM proceedings, \LaTeX, text tagging}
%\input{open}

\bibliographystyle{ACM-Reference-Format}
\bibliography{biblio}

\appendix
%\clearpage
% flatex input: [theoretical-results.tex]
%!TEX root = main.tex

\section{A formal definition of G-CORE}
\label{theory}

%\textcolor{blue}{\bf{george to marcelo:} we still need to add GRAPH and GRAPH VIEW}

In this section we formally present the semantics of \gcore{} queries.  For ease of presentation, we focus on a simplified syntax equal in expressive power to the full syntax of the language.

% flatex input: [match.tex]
%!TEX root = main.tex

%\subsection{The general structure of a G-CORE query}
%\label{sec:full-grammar} 

\subsection{Basic notions}
\label{sec:basic-notions}

$\gcore$ queries are recursively defined by the top-down grammar presented in
Section~\ref{sec:formal}. In this section we define in detail the different
components of the $\gcore$ grammar. But before doing so, it is important to
introduce some basic notions that are used in the definition of their
semantics.

Recall the domains used by the $\gcgraph$ data model as defined in Section~\ref{sec:datamodel}. 
Let $\bL$ be an infinite set of label names for nodes, edges and paths, $\bK$ an infinite set of property names and $\bV$ an infinite set of literals (i.e. actual values like integers). 

%\vspace*{-1mm}
\mypar{\underline{Paths conforming to regular expressions}}
In graph query languages, one is typically interested in checking if two nodes are linked by a path whose label satisfies a regular expression $r$, and computing 
one (or more) of such paths if needed. Some graph query languages. e.g., Cypher~9~\cite{Cypher18}, define the semantics of path expressions based on {\em simple paths}
only (those without repetition of nodes and/or edges), which is known to easily lead to intractability in data complexity~\cite{MW95}. For this reason, 
in G-CORE we follow a long tradition of graph query languages introduced in the last 30 years and 
define the semantics of path expressions based on arbitrary paths (see, e.g.,~\cite{AABHRV16}). In addition, for the problem of computing  
a path from node $u$ to node $v$ that satisfies a given regular expression $r$, we choose to compute the {\em shortest} such a path according 
to a fixed lexicographical order on nodes.\footnote{We acknowledge that using a fixed lexicographical order could be too restrictive when choosing a single path, so a system implementing $\gcore$ could use a different criterion based, for instance, on whether it can be evaluated more efficiently.}
The reason for this is that checking for the existence of an arbitrary path in a $\gcgraph$ $G$ from $u$ to $v$ 
that conforms to a regular expression $r$, and computing
the shortest path that witnesses this fact, can be done in polynomial time by applying standard automata techniques in combination 
with depth-first search. 
We define the notion of (shortest) paths conforming to regular expressions below.  

We start by defining the notion of regular expression used in \gcore. A regular expression $r$ is specified by the grammar:
\begin{eqnarray*}
r & ::= & \und \ \mid \ \ell \ \mid \ \ell^-   \ \mid \ \nl{\ell} \ \mid \  (r + r) \ \mid \ (r r) \ \mid \ (r)^*,
\end{eqnarray*}
where $\ell \in \bL$. Intuitively, an expression of the form either $\ell$ or $\ell^-$, with $\ell \in \bL$, refers to an edge label, while $\nl{\ell}$ refers to a node label. The expression $\und$ is used as a wildcard that stands for ``any label''. Notice that under this definition, the alphabet of every regular expression is a finite subset of $\{ \und \} \cup \bL \cup \{ \ell^- \mid \ell \in \bL  \} \cup \{ \nl{\ell} \mid \ell \in \bL  \}$.

Let  $G = (N, E, P, \rho, \delta, \lambda, \sigma)$ be a $\gcgraph$ and $L = [a_1, e_1, a_2, \ldots, a_n, e_n, a_{n+1}]$ be a path over $G$. 
%From now on, we say that $a_1$ and $a_{n+1}$ are the starting and ending nodes of $L$. 
Then given a regular expression $r$, we say that $L$ is a path from $a_1$ to $a_{n+1}$ conforming to $r$ if there exists a string $u_1 v_2 u_2 \ldots u_n v_{n+1},  u_{n+1}$ in the regular language defined by $r$ such that:
\begin{itemize}
%\item For every $i \in \{1, \ldots, n+1\}$, either $u_i = \und$ or $u_i = \nl{\ell}$ for $\ell \in \bL$.

%\item For every $j \in \{1, \ldots, n\}$, either $v_j = \und$ or $v_j = \ell$ or $v_j = \ell^-$, for $\ell \in \bL$.

\item For each $i \in \{1, \ldots, n+1\}$, either $u_i = \und$ or $u_i = \nl{\ell}$ for some $\ell \in \lambda(a_i)$.

\item For each $j \in \{1, \ldots, n\}$, either  

\begin{itemize} 

\item $v_j = \und$ , or

\item $\rho(e_j) = (a_{j},a_{j+1})$ and $v_j = \ell$ for some $\ell \in \lambda(e_j)$, or

%%with $v_j = \ell$, it is the case that $\ell \in \lambda(e_j)$ and $\rho(e_j) = (a_{j},a_{j+1})$.
%%
%%\item For each $j \in \{1, \ldots, n\}$ with $v_j = \ell^-$, it is the case that $\ell \in \lambda(e_j)$ and $\rho(e_j) = (a_{j+1},a_{j})$.
%\end{itemize}
%We formalize next the notion of shortest path we use in the paper. 

\item $\rho(e_j) = (a_{j+1},a_{j})$ and $v_j = \ell^-$ for some $\ell \in \lambda(e_j)$.
\end{itemize} 

\end{itemize} 

The length of a path $L = [a_1, e_1, a_2, \ldots, a_n, e_n, a_{n+1}]$, written $\length(L)$, is $n$. Then $L$ is a {\em shortest path} from a node $a$ to a node $b$ conforming to a regular expression $r$, if for every path $L'$ from $a$ to $b$ that conforms to $r$, it holds that $\length(L) \leq \length(L')$.

%\vspace*{-1mm}
\mypar{\underline{Bindings}}
From now on, we assume that $\cN$, $\cE$ and $\cP$ are countably 
infinite sets of node, edge and path variables, respectively, which are pairwise disjoint.  
%In order to define the semantics of basic graph patterns, we need to introduce some terminology. 
Let $G = (N, E, P, \rho, \delta, \lambda, \sigma)$ be a $\gcgraph$. A {\em binding} $\mu$ over $G$ is a partial 
function $\mu : (\cN \cup \cE \cup \cP) \to (N \cup E \cup P)$ such that $\mu(x) \in N$ if $x \in \cN$, $\mu(y) \in E$ if $y \in \cE$, and $\mu(z) \in P$ if $z \in \cP$. 
The domain of a binding $\mu$ is denoted by $\dom{\mu}$, and it is assumed to be finite. 
Two bindings $\mu_1$ and $\mu_2$ are said to be {\em compatible}, denoted by $\mu_1 \sim \mu_2$, if for every variable $x \in \dom{\mu_1} \cap \dom{\mu_2}$, it holds that $\mu_1(x) = \mu_2(x)$. Notice that if $\mu_1$ and $\mu_2$ are compatible bindings, then $\mu_1 \cup \mu_2$ is a well-defined function.

Let $\Omega_1$, $\Omega_2$ be finite sets of bindings over $G$. 
The following four basic operations will be extensively used in this article:
\begin{align*}
\Omega_1 \cup \Omega_2    = &\ \{ \mu \mid \mu \in \Omega_1 \text{ or } \mu \in \Omega_2 \},\\
\Omega_1 \Join \Omega_2   = &\ \{ \mu_1 \cup \mu_2 \mid \mu_1 \in \Omega_1, \mu_2 \in \Omega_2 \text{ and } \mu_1 \sim \mu_2\},\\
%\Omega_1 \ltimes \Omega_2 = &\ \{ \mu_1 \mid \mu_1 \in \Omega_1 \text{ and there is}\\ 
%& \hspace{50.4pt} \mu_2 \in \Omega_2 \text{ such that } \mu_1 \sim \mu_2\},\\
\Omega_1 \ltimes \Omega_2 = &\ \{ \mu_1 \mid \mu_1 \in \Omega_1, \mu_2 \in \Omega_2 \text{ and } \mu_1 \sim \mu_2\},\\
\Omega_1 \smallsetminus \Omega_2 = &\ \{ \mu_1 \mid \mu_1 \in \Omega_1 \text{ and } \nexists\mu_2\!\in\!\Omega_2, \mu_1 \sim \mu_2\},\\
\Omega_1 \leftouterjoin \Omega_2 = &\ (\Omega_1 \Join \Omega_2) \cup (\Omega_1 \smallsetminus \Omega_2). 
\end{align*}

Given a countably infinite set of value variables $\cV$ disjoint from $\cN$, $\cE$ and $\cP$, it is straightforward to 
extend our definition of bindings (and our presentation of the semantics of \gcore) to capture assignments of literals to variables, i.e., 
such an extended binding would be a partial function
$\mu : (\cN \cup \cE \cup \cP \cup \cV) \to (N \cup E \cup P \cup \bV)$ such that $\mu(x)\in \bV$ if $x\in\mathcal{V}$.
For ease of presentation, however, we do not further consider such extended bindings in the sequel.

% flatex input: [where.tex]
%!TEX root = main.tex

%\vspace*{-1mm}
\mypar{\underline{Expressions}}\label{sec:expressions}
We next define expressions $\expr$ according to the following grammar:  
\begin{align*}
\expr ::= &\ x \ \mid\ 
             \propaccess{x}{k}  \ \mid\ 
             \labelcheck{x}{\ell}  \ \mid\ 
             \diamond\expr  \ \mid\ 
             \expr\odot\expr  \ \mid\ 
        f(\expr,\expr,\ldots)  \ \mid\ 
             \Sigma(\expr)  \ \mid\ 
             \Exists{}\ q,
\end{align*}
where $x$ is a variable, $k\in\bK$ is property key, $\ell\in\bL$ is a label, $\diamond$ is a unary operator, $\odot$ is a binary operator, $f$ is a built-in function whose value only depends on its inputs and $\Sigma$ is an aggregation function.
Unary operators are boolean negation \Not{}, arithmetic negation $-1$, etc.
Binary operators are comparisons such as $=,\neq,<,\leq,\Subsetof,\In,\ldots$, 
boolean operations such as \Andkw{} and \Orkw{}, arithmetic operations on numbers and strings such as $+,-,*,/,\ldots$, etc.
Built-in functions include the standard ones for type casting, string, date and collection handling (e.g. \Size) known from other query languages, 
as well as graph-specific function such as \Labels{}, which returns the
set of labels, and \Nodes{}, and \Edges{}, which return the list of nodes and edges of $o$, respectively, when applied to an object identifier $o$.
Aggregation functions include the standard ones inherited from relational query languages, \Count, \Min, \Max, \Sum, \Avg, etc. and \Collect{} to collect all values of the group into a collection.

The semantics of expressions inherited from other query languages is defined in analogy with them, so we do not repeat it here.
We denote the evaluation of any such an expression \expr\ over a $\gcgraph$ $G = (N, E, P, \rho, \delta, \lambda, \sigma)$ and 
a set of bindings $\Omega$ 
as $\eval{\expr}{\Omega,G}$. 
In any such a case, $\eval{\expr}{\Omega,G} \in N \cup E \cup P \cup \bL \cup \bV$ (recall that $\bV$ includes truth values $\bot$ and $\top$).
The semantics of the \gcore{} specific expressions, on the other hand, is defined as follows (where we denote by $\mu$ the singleton binding set $\set{\mu}$):

\begin{itemize}
    \item If $\expr$ is a variable $x$, then $\eval{\expr}{\mu,G}=\eval{x}{\mu,G} = \mu(x)$.
	\item If $\expr = \propaccess{x}{k}$, then $\eval{\expr}{\mu,G}=\sigma(\eval{x}{\mu,G},k) = \sigma(\mu(x),k)$.
	\item If $\expr = \labelcheck{x}{\ell}$, then $\eval{\expr}{\mu,G} = \top$ iff $\ell \in \lambda(\eval{x}{\mu,G})$ iff $\ell \in \lambda(\mu(x))$. 
	\item If $\expr = \Labels(o)$, then $\eval{\expr}{\mu,G}=\lambda(\eval{o}{\mu,G})$.
	\item If $\expr = \Nodes(o)$, then $\eval{\expr}{\mu,G} = {\tt nodes}(\eval{o}{\mu,G})$. 
	%[a_1, \ldots, a_{n+1}]$ with $\delta(\eval{x}{\mu,G}) = [a_1, e_1, a_2, \ldots, e_n, a_{n+1}]$.
	\item If $\expr = \Edges(o)$, then $\eval{\expr}{\mu,G}={\tt edges}(\eval{o}{\mu,G})$. 
	%[e_1, \ldots, e_n]$ with $\delta(\eval{x}{\mu,G}) = [a_1, e_1, a_2, \ldots, e_n, a_{n+1}]$.
	\item If $\expr = \Exists \,\, q$, then $\eval{\expr}{\mu,G}= \top$ iff $(N\neq\emptyset)$, assuming that 
	$\eval{q}{\mu,G}=(N, E, P, \rho, \delta, \lambda, \sigma)$.
\end{itemize}

%  A scalar value is a variable, a constant, or the result of some aggregation function 
% over some collection. Such aggregation functions include the standard ones inherited from relational query languages, 
% e.g., {\tt MIN}, {\tt MAX}, {\tt SUM}, {\tt AVG}, etc., but also others that are of particular interest when dealing with graph data. 
% An example is the function {\tt LENGTH} that returns the length of a collection of objects (in particular, the length of a path). 

%  Such collections of values
% are obtained from an object identifier $o$ by applying operations {\tt LABELS}, {\tt NODES}, and {\tt EDGES}, which return the collection 
% of labels, nodes, and edges in $o$, respectively. 

% flatex input end: [where.tex]
 
%& \hspace{50.4pt} \mu_2 \in \Omega_2 \text{ such that } \mu_1 \sim \mu_2\},\\

\subsection{The MATCH clause }\label{sec:match}

\mypar{\underline{Basic graph patterns}}
A {\em basic graph pattern} is specified by the following grammar:
%\vspace*{-1mm}
\begin{align*}
%\basic graph pattern &::=& (x) \ \mid \ \ep \ \mid \ \pp\\
\bgp ::= &\ \np \ \mid \ \ep \ \mid \ \pp\\
\np  ::= &\ (x)\\
\ep  ::= &\ \edgepattern{x}{~z~}{y}\\ 
% \; \ \mid \ \; \uedgepattern{x}{~z~}{y}\\
\pp  ::= &\ \pathpattern{x}{w}{r}{y} \ \mid \ \wpathpattern{x}{w}{r}{y}
% \ \mid\\ && \upathpattern{x}{w}{r}{y} \ \mid \ \wupathpattern{x}{w}{r}{y}
\end{align*}
where $x,y \in \cN$, $z \in \cE$, $w \in \cP$, and $r$ is a regular expression.

% 
%A $\FGP$ evaluated on $G$ returns a set of mappings.
%A {\em mapping} is a partial function $\mu:
%\mathcal{N}\cup \mathcal{E} \cup \mathcal{P} \to N \cup E \cup P$ with finite domain
%which maps elements of $\mathcal{N}$, $\mathcal{E}$, and $\mathcal{P}$ to elements of $N$, $E$, and $P$, respectively. 

Let $\alpha$ be a basic graph pattern. The evaluation of $\alpha$ over a $\gcgraph$ $G = (N, E, P, \rho, \delta, \lambda, \sigma)$, denoted by 
$\eval{\alpha}{G}$, is inductively defined: % as follows.
\begin{itemize}
    \item If $\alpha$ is a node pattern $(x)$, then
        $\eval{\alpha}{G} = \{ \mu \mid  \dom{\mu} = \{x\} \text{ and } \mu(x)\in N\}$. 
    \item If $\alpha$ is an edge pattern $\edgepattern{x}{z}{y}$, then 
       % \begin{multline*} 
        $\eval{\alpha}{G} = \{ \mu \mid \dom{\mu} = \{x,y,z\}, \, \mu(x),\mu(y) \in N, \mu(z) \in E \text{ and } \rho(\mu(z)) = (\mu(x), \mu(y))\}$.
        %\end{multline*} 
%    \item If $\basic graph pattern$ is an edge pattern \uedgepattern{x}{z}{y}, then 
%        $\eval{\basic graph pattern}{G} = \eval{\edgepattern{x}{z}{y}}{G} \cup \eval{\edgepattern{y}{z}{x}}{G}$.
    \item If $\alpha$ is a path pattern \pathpattern{x}{w}{r}{y}, then 
%        \begin{multline*}
        $\eval{\alpha}{G} = \{\mu \mid \dom{\mu} = \{x,y,w\}$, $\mu(x), \mu(y) \in N$, $\mu(w) \in P$ and
        $\delta(\mu(w))$ is the shortest path from $\mu(x)$ to $\mu(y)$ that conforms to $r\}.$
%        \end{multline*}
            \item If $\alpha$ is a path pattern \wpathpattern{x}{w}{r}{y}, then 
%        \begin{align*}
       % \begin{multline*} 
        $\eval{\alpha}{G} = \{\mu \mid\ \dom{\mu} = \{x,y,w\}, \mu(x), \mu(y) \in N$, and  
        $\mu(w)$ is a fresh path identifier (i.e., $\mu(w) \not\in P$) associated  
        to the shortest path $L$ from $\mu(x)$ to $\mu(y)$ over $G$ that conforms to 
        $r \}$.
        %\end{multline*} 
%        \end{align*}
%        \item If $\basic graph pattern$ is a path pattern \upathpattern{x}{z}{r}{y}, then 
%$\eval{\basic graph pattern}{G} = \eval{\pathpattern{x}{z}{r}{y}}{G} \cup \eval{\pathpattern{y}{z}{r}{x}}{G}$.
%        \item If $\basic graph pattern$ is a path pattern \wupathpattern{x}{z}{r}{y}, then 
%$\eval{f}{G} = \eval{\wpathpattern{x}{z}{r}{y}}{G} \cup \eval{\wpathpattern{y}{z}{r}{x}}{G}$.
%        
%    \noop{
%    \item If $f$ is $\sigma_C(g)$ for some $g\in \FGP$, then $\eval{f}{G}f(G) = \{\mu \in g(G) \mid \mu \textnormal{ satisfies } C\}$.
%    }
%    %\item If $f$ is $(f_1, f_2)$ for some full graph patterns $f_1$ and $f_2$, then $\eval{f}{G} = \eval{f_1}{G} \Join \eval{f_2}{G}$.        
\end{itemize}

%\vspace*{-1mm}
\mypar{\underline{Basic graph patterns with location and full graph patterns}}\label{sec-fgp}
Basic graph patterns can be evaluated over different graphs in \gcore. More specifically, we use 
an $\on$ operator to specify the location where a basic graph pattern has to be evaluated:
\begin{align*}
\bgpl ::= &\ \bgp \ \mid \\
          &\ \bgp \ \on \ \loc\\
\loc  ::= &\ \gid \ \mid \ \fgquery , 
\end{align*}
where $\gid$ is a graph identifier and $\fgquery$ is as defined in the G-CORE grammar given in Section~\ref{sec:formal}. 
For reasons that will become clear later, we can always assume the evaluation $\eval{\alpha}{G}$ of a full graph query $\alpha$ over a $\gcgraph$
$G$ to be another $\gcgraph$ $H$. 

To define the semantics of basic graph patterns with location, we assume given a function $\graph$ that associates an actual graph, denoted by 
$\graph(\gid)$, to every graph identifier $\gid$. 
Then given a basic graph pattern with location $\beta$, the evaluation of $\beta$ over a $\gcgraph$ $G$, denoted by 
$\eval{\beta}{G}$, is defined as follows.
\begin{itemize}
\item If $\beta$ is $\alpha \ \on \ \gid$, where $\alpha$ is a basic graph pattern and $\gid$ is a graph identifier, then $\eval{\beta}{G} = \eval{\alpha}{\graph(\gid)}$.

\item If $\beta$ is $\alpha \ \on \ Q$, where $\alpha$ is a basic graph pattern, $Q$ is a full graph query, 
and $H = \eval{Q}{G}$, then $\eval{\beta}{G} = \eval{\alpha}{H}$.
\end{itemize}
Finally, a full graph pattern is defined as a sequence of basic graph patterns with locations separated by comma:
\begin{align*}
\fgp  ::= &\ \bgpl \ \mid \\
          &\ \bgpl,\ \fgp
\end{align*}
To define the evaluation of such pattern $\gamma$ over a $\gcgraph$ $G$, denoted by $\eval{\gamma}{G}$, we only need to consider the following recursive rule:
\begin{itemize}
\item If $\gamma$ is a $\beta, \gamma'$, where $\beta$ is a basic graph pattern with location and $\gamma'$ is a full graph pattern, then $\eval{\gamma}{G} = \eval{\beta}{G} \Join \eval{\gamma'}{G}$.  
\end{itemize}

%\vspace*{-1mm}
\mypar{\underline{Conditions on binding tables}} 
\gcore\ includes a \where{} clause that allows to filter bindings according to some boolean 
{\em conditions}. It is important to mention that this clause cannot be used to generate new bindings, 
it can only be used to filter the bindings that are produced by a \match{} clause.

A \where{} clause can include conditions of the form $\kw{Exists} \ Q$, where $Q$ is a sub-query. Such a query evaluates to true if 
the evaluation of $Q$ is a non-empty $\gcgraph$, and it evaluates to false otherwise. It is important to 
consider that $Q$ can share some variables with the outer query, which are considered as parameters as in the case of 
correlated subqueries in SQL. Thus, in order to define the semantics of $\Exists \ Q$, it is necessary to extend the definition of the semantics of full graph patterns to consider some previously computed bindings. More precisely, given a full graph pattern 
$\gamma$ and a set $\Omega$ of bindings, the evaluation of $\gamma$ over $G$ given $\Omega$, denoted by 
$\eval{\gamma}{\Omega,G}$, is defined as
%\begin{eqnarray*}
$\eval{\gamma}{\Omega,G} = \eval{\gamma}{G} \ltimes \Omega$. 
%\end{eqnarray*}
Notice that if $\Omega = \{ \mu_\emptyset \}$, where $\mu_\emptyset$ is the binding with empty domain, then $\eval{\gamma}{\Omega,G} = \eval{\gamma}{G}$. 
%Thus, this definition of the semantics of full graph patterns is an extension of the definition given in Section \ref{sec-fgp}.

We now have the necessary ingredients to define the semantics of MATCH clauses with conditions. The syntax of such expressions is given by the following grammar:
\begin{align*}
\mclause ::=  \, & \match \ \fgp \ \mid \\
& \kw{Match} \ \fgp \ \where \ \bc,
\end{align*}
where $\bc$ is a Boolean-valued expression, that is, an expression that evaluates to $\top$ or $\bot$ 
(expressions are formally defined in Section~\ref{sec:expressions}). 
To define the evaluation of a MATCH clause over a $\gcgraph$ $G$ given a set of $\Omega$ of bindings, we only need to consider the following rule:
\begin{itemize}
\item If $\gamma$ is a full graph pattern and $\expr$ is a Boolean-valued expression, then $\eval{\mw{\gamma}{\expr}}{\Omega,G} $ = $\{ \mu \in \eval{\gamma}{\Omega,G} \mid  \eval{\expr}{\mu,G} = \top \}$.
\end{itemize}

%$\mw{\gamma}{\expr}$, where $\gamma$ is a full graph pattern and $\expr$ is a Boolean-valued expression, that is, an expression that evaluates to 0 or 1, which represents true and false, respectively (expressions are formally defined in Section \ref{sec:expressions}). More precisely, given a $\gcgraph$ $G$ and a set $\Omega$ of bindings, the evaluation of $\mw{\gamma}{\expr}$ over $G$ given $\Omega$, denoted by $\eval{\mw{\gamma}{\expr}}{G,\Omega}$, is defined as:
%%we next define the semantics of \Where{} clause ``$\mw{f}{\expr}$'' on a $\gcgraph$ $G$.   
%%Intuitively, the result set will consist of all mappings in $\eval{f}{G}$ which satisfy $\expr$.   The only subtlety is to properly handle correlations between $f$ and $\expr$, through shared variables.  We handle this by evaluating $\expr$ in the context of mappings of $f$ in $G$.
%\begin{eqnarray*}
%\eval{\mw{\gamma}{\expr}}{G,\Omega} & = & \{ \mu \in \eval{\gamma}{G,\Omega} \mid  \eval{\expr}{G,\{\mu\}} = 1 \}.
%\end{eqnarray*}

\mypar{\underline{Optional full graph patterns}} 
In its most general version, full graph patterns can also use an \optional{} operator to include optional information if present. 
Thus, the general syntax of \match{} clauses is specified by the following grammar:
\begin{align*}
&\fgpc ::= \ \fgp \ \mid \\
& \hspace{50pt} \fgp \ \where \ \bc\\
&\mclause ::= \ \match \ \fgpc \ \mid\\
& \hspace{50pt} \match \ \fgpc \ \oclause
\end{align*}
where $\oclause$ is given by the following grammar:
\begin{align*}
&\oclause ::= \ \optional \ \fgpc \ \mid \\ 
& \hspace{50pt} \optional \ \fgpc \ \oclause
\end{align*}
The semantics of the \optional{} operator is defined by using the operator $\leftouterjoin$ on sets of bindings. More precisely, given a sequence of full graph patterns $\gamma$, $\gamma_1$, $\ldots$, $\gamma_k$, where $k \geq 1$, and a sequence of Boolean conditions $\expr$, $\expr_1$, $\ldots$, $\expr_k$, we have that:
\begin{align*}
\llbracket&\mw{\gamma}{\expr} \\ 
&\ow{\gamma_1}{\expr_1} \cdots\ \ow{\gamma_k}{\expr_k}\rrbracket_{G,\Omega} =\ \\
&((\cdots((\eval{\mw{\gamma}{\expr}}{G,\Omega} \ \leftouterjoin \ \eval{\mw{\gamma_1}{\expr_1}}{G,\Omega}) \\ 
&\hspace{24pt} \leftouterjoin \ \cdots) \cdots) \ \leftouterjoin \ \eval{\mw{\gamma_k}{\expr_k}}{G,\Omega}). 
\end{align*}

%\vspace*{-1mm}
\mypar{\underline{Example}} 
Consider $\mw{\gamma}{\expr}$ where
%\vspace*{-2mm}
\begin{align*}
    \gamma =&\ \edgepattern{x}{~\elt{locatedIn}~}{w}, \ \edgepattern{y}{~\elt{locatedIn}~}{w}, \ \pathpattern{x}{z}{(\elt{knows} + \elt{knows}^-)^*}{y} \\
    \expr  =&\ w.\elt{name} = Houston.
\end{align*}
On the $\gcgraph$ $G = (N, E, P, \rho, \delta, \lambda, \sigma)$ given in Figure \ref{fig:small-social-network} and formalized in Section \ref{sec:datamodel}, 
we have that
\begin{align*} 
    \eval{\gamma}{G}  =&\ \eval{\edgepattern{x}{~\elt{locatedIn}~}{w}}{G} \Join \eval{\edgepattern{y}{~\elt{locatedIn}~}{w}}{G} \\
                       &\ \Join \eval{\pathpattern{x}{z}{(\elt{knows} + \elt{knows}^-)^*}{y}}{G} \\
                      =&\ \{\{x\mapsto 105, w \mapsto 106\}, \{x\mapsto 102, w \mapsto 106\}\}  \\ 
                       &\ \Join \{\{y\mapsto 102, w \mapsto 106\}, \{y\mapsto 105, w \mapsto 106\}\} \\ 
                       &\ \Join \{\{z\mapsto 301, x \mapsto 105, y \mapsto 102\}\} \\
                      =&\ \{ \{x\mapsto 105, y\mapsto 102, w \mapsto 106\}, \{x\mapsto 105, y\mapsto 105, w \mapsto 106\}, \\ 
                       &\     \{x\mapsto 102, y\mapsto 102,w \mapsto 106\}, \{x\mapsto 102, y\mapsto 105,w \mapsto 106\} \}  \\ 
                       &\ \Join \{\{z\mapsto 301, x \mapsto 105, y \mapsto 102\}\} \\
                      =&\ \{ \{x\mapsto 105, y\mapsto 102, w \mapsto 106, z\mapsto 301\} \}
\end{align*}
and, letting $\mu = \{x\mapsto 105, y\mapsto 102, w \mapsto 106, z\mapsto 301\}\in \eval{\gamma}{G}$, we then have that
%\vspace*{-1mm}
\begin{align*} 
\eval{\expr}{\mu, G} =&\ \eval{w.\elt{name} = Houston}{\mu, G} = \top.
\end{align*}
Therefore, 
$\eval{\mw{\gamma}{\expr}}{G} = 
\{ \{x\mapsto 105, y\mapsto 102, w \mapsto 106, z\mapsto 301\} \}.$

% flatex input end: [match.tex]

%\textcolor{blue}{\bf{george to marcelo:} we still need to add GRAPH and GRAPH VIEW}
%\input{where}
% flatex input: [construct.tex]
%!TEX root = main.tex

\newcommand{\copypattern}[2]{\brk{+#1 = #2}}
\newcommand{\setprop}[3]{\brk{+#1.#2 = #3}}
\newcommand{\removeprop}[2]{\brk{-#1.#2}}
\newcommand{\setlabel}[2]{\brk{+#1:#2}}
\newcommand{\removelabel}[2]{\brk{-#1:#2}}

\newcommand{\constructvars}{\ensuremath{\mathcal{C}}}
\newcommand{\FGCP}{full graph construction patterns}
\newcommand{\nodecpattern}[3]{\ensuremath{(#1\;\text{\Group}\;#2; #3)}}
\newcommand{\edgecpattern}[5]{\ensuremath{#1\xrightarrow{#2\;\text{\Group}\;#3; #5}#4}}
\newcommand{\pathcpattern}[5]{\ensuremath{#1\;\xrsquigarrow{#3#2; #5}\;#4}}
\newcommand{\conditionalcpattern}[2]{\ensuremath{#1\;\when\;#2}}
\newcommand{\unioncpattern}[2]{\ensuremath{#1,#2}}

\newcommand{\boundvars}{B}
\newcommand{\grpvars}{\ensuremath{\Gamma}}

\newcommand{\newobj}[1]{\ensuremath{\op{new}(#1)}}
\newcommand{\name}[1]{\ensuremath{\op{name}(#1)}}

\newcommand{\grp}[1]{\ensuremath{\op{grp}(#1)}}

\subsection{The CONSTRUCT clause}\label{sec:construct}

The definition of the \construct{} clause assumes finite set of variables $\boundvars\subset\cN\cup\cE\cup\cP\cup\cV$.
The set contains the variables declared in all full graph patterns of the \match{} clause. 
$\boundvars$ is a syntactical property of a given query.
We use $\emptygraph$ to denote an empty graph.

%\subsubsection{Full Graph Construction Pattern}
%with respect to a set of mappings $\Omega$

\newcommand{\fcp}{\text{\it fullConstruct}}
\newcommand{\bcp}{\text{\it basicConstruct}}
\newcommand{\ocps}{\text{\it objectConstructsList}}
\newcommand{\ocp}{\text{\it objectConstruct}}
\newcommand{\ccp}{\text{\it combiConstruct}}
\newcommand{\ncp}{\text{\it nodeContruct}}
\newcommand{\rcp}{\text{\it relationshipContruct}}
\newcommand{\ecp}{\text{\it edgeContruct}}
\newcommand{\pcp}{\text{\it pathContruct}}
\newcommand{\wcp}{\text{\it conditionalContruct}}

\newcommand{\asagraph}{\text{graph}}
\newcommand{\asbindings}{\text{bindings}}
\newcommand{\aslabels}{\text{labels}}
\newcommand{\asproperties}{\text{properties}}

\newcommand{\restrict}[2]{\ensuremath{#1\!\mid_{#2}}}

%\vspace*{-1mm}
\mypar{\underline{Syntax of Basic Construct Patterns}}
The {\em basic constructs} are specified by the following grammar:
%\vspace*{-2mm}
\begin{align*}
\bcp  ::= &\ \ocps \ \mid \\
          &\ \conditionalcpattern{\ocps}{\expr} \\
\ocps ::= &\ \gid \ \mid \ \ocp \ \mid \\
          &\ \ocp \ \ocps 
\end{align*}
where $\gid$ is a graph identifier.

{\em Object constructs} with respect to the set of variables $\boundvars$ are specified by the following grammar:
%\vspace*{-2mm}
\begin{align*}
\ocp ::= &\ \ncp \ \mid \ \rcp \\
\rcp ::= &\ \ecp \ \mid \ \pcp \\
\ncp ::= &\ \nodecpattern{x}{\grpvars{}}{S} \\
\ecp ::= &\ \edgecpattern{x}{z}{\grpvars{}}{y}{S} \\
\pcp ::= &\ \pathcpattern{x}{u}{@}{y}{S} 
\end{align*}
where $x, y\in \mathcal{N}$, $z\in \mathcal{E}$, $u \in \mathcal{P}$, $\grpvars{} \subseteq \boundvars$ is a grouping sets,
$S$ is a set of assignments,
and $o$ is the optional symbol $@$ (i.e., can appear or can be left out).
Each object construct pattern has a {\em construct variable}, which is unique within the basic construct.
For a node construct \nodecpattern{x}{\grpvars{}}{S}, an edge construct \edgecpattern{x}{z}{\grpvars{}}{y}{S}, and a path construct \pathcpattern{x}{u}{@}{y}{S}, the construct variable is $x$, $z$, and $u$, respectively.
In situation of ambiguity, we refer to $\grpvars{}$ and $S$ of a specific object construct $g$ by subscripting $\grpvars{}$ and $S$ with $g$'s construct variable.

We further constrain the syntax of a basic construct $b$ used in the context of a basic graph query, as follows.
\begin{itemize}
    \item In each node construct appearing in $b$, 
    if $x\in\boundvars$ then $\grpvars{}=\set{x}$.
    \item In each edge construct \edgecpattern{x}{z}{\grpvars{z}}{y}{S_z} appearing in $b$,
        it holds that 
        \begin{itemize}
            \item node construct \nodecpattern{x}{\grpvars{x}}{S_x}\ and \nodecpattern{y}{\grpvars{y}}{S_y} must appearing in $b$; %%% NOTE: we need this otherwise the WHEN allow to construct dangling edges!!!
            \item $\grpvars{x} \cup \grpvars{y} \cup \set{x,y} \subseteq \grpvars{z}$; and
            \item if $z\in\boundvars$, then $x,y\in\boundvars$.
        \end{itemize}
        Furthermore, if $z\in\boundvars$, then $\grpvars{z}=\set{z}\cup\grpvars{x}\cup\grpvars{y}$.

    \item In each path construct \pathcpattern{x}{u}{o}{y}{S} appearing in $b$, 
        it must hold that a path pattern \pathpattern{x}{u}{r}{y} appears in the match clause, i.e. $x,y,u\in \boundvars$.
        By definition the grouping set for a path construct is $\grpvars{}=\set{x,y,u}$.
\end{itemize}

The conditions ensure, that all object constructs in a basic construct form a valid directed graph and, hence, the use of the \when{} sub-clause can not cause the creation of dangling edges in the resulting $\gcgraph$.
%For a basic construct $b$, let $N_b$ be the set of construct variables of all node constructs in $b$.
%Each relationship construct $r$ additionally refers to a pair of variables $x$ and $y$.
%Let $R_b$ be the set of all pairs referred to by relationship constructs in $b$.
%Finally, the syntax of basic constructs requires that $(N_b, R_b)$ is a valid directed graph.
%%% NOTE: we need this otherwise the WHEN allow to construct dangling edges!!!

\mypar{\underline{Semantics of Basic Construct Patterns}}
The semantics of any basic construct $b$ with respect to a set of bindings $\Omega$ evaluated on $G$, is a pair $\brk{\eval{b}{\Omega,G}^{\asagraph},\eval{b}{\Omega,G}^{\asbindings}}$, where the $\eval{b}{\Omega,G}^{\asagraph}$ is a $\gcgraph$ and $\eval{b}{\Omega,G}^{\asbindings}$ is a set of bindings.
In case where $b$ is a graph identifier \gid, the $\gcgraph$ resulting from the basic construct pattern is the actual graph associated with \gid, i.e. $\eval{\gid}{\Omega,G}^{\asagraph} = \graph(\gid)$ and $\eval{\gid}{\Omega,G}^{\asbindings} = \emptybinding$.

For the remaining cases, without loss of generality, we consider a basic construct $b$ to be \conditionalcpattern{L}{\expr}, where $L$ is the set of all object constructs in $b$.
We assume $\expr=\top$ if the \when{} sub-clause is not given in the query.
It should be mentioned that the \when{} sub-clause is not syntactic sugar. 
Since basic queries form the scope for variables, the \when{} sub-clause can not be rewritten to a full query with \Union{}.

The semantics of a basic construct is defined as:
%\vspace*{-1mm}
\begin{align*}
\eval{\conditionalcpattern{L}{\expr}}{\Omega,G}^{\asagraph} =&\ 
\begin{cases}
    \eval{L}{\Omega,G}^{\asagraph} & \text{if } \eval{\expr}{\Omega^\expr,G}=\top \\
    \emptygraph & \text{otherwise}, 
\end{cases} \\
\eval{\conditionalcpattern{L}{\expr}}{\Omega,G}^{\asbindings} =&\ 
\begin{cases}
    \eval{L}{\Omega,G}^{\asbindings} & \text{if } \eval{\expr}{\Omega^\expr,G}=\top \\
    \emptybinding & \text{otherwise}. 
\end{cases}
\end{align*}
where $\Omega^\expr = \Omega \Join \eval{\conditionalcpattern{L}{\expr}}{\Omega,G}^{\asbindings}$. Note, that $\Omega^\expr$ ensures that the expression $\expr$ in a \when{} sub-clause can use variables appearing the \match{} clause ($\boundvars$) as well as construct variables of the constructs.

%%% NOTE: this actually not absolutely correct, sind SET/REMOVE are evaluated be the WHEN, so that S has to become part of the pattern again

For the semantics of $L$, let $L_N$ be the set of all node constructs in $L$ and $L_R$ the set of all relationship constructs (edge and path constructs) in $L$.
We define the semantics of $L$ such that the \gcgraph{}s and the sets of bindings resulting from evaluation of all object constructs in $L$ are united and joined, respectively.
The union of graphs is defined as in Section~\ref{sec:fullquery}.
Further, we ensure that the evaluation of all relationship constructs ($L_R$) is based on the set of bindings resulting from the evaluation of all node constructs ($L_N$), so new edges connect to new nodes.
The semantics of $L$ is:
%\vspace*{-1mm}
\[\begin{array}{l@{\ }c@{\ }l@{\ }c@{\ }l}
\eval{L}{\Omega,G}^{\asagraph}   &=& G_N      &\cup & \brk{\bigcup_{g\in L_R} \eval{g}{\Omega_N,G}^{\asagraph}} \\
\eval{L}{\Omega,G}^{\asbindings} &=& \Omega_N &\Join& \brk{  \Join_{g\in L_R} \eval{g}{\Omega_N,G}^{\asbindings}}
\end{array}\]
with: 
%\vspace*{-5.2mm}
\[\begin{array}{l@{\ }c@{\ }l@{\ }c@{\ }l@{\ }c@{\ }l}
    G_N      &=&        &     &\eval{L_N}{\Omega,G}^{\asagraph}   &=& \bigcup_{g\in L_N} \eval{g}{\Omega,G}^{\asagraph} \\
    \Omega_N &=& \Omega &\Join& \eval{L_N}{\Omega,G}^{\asbindings} &=& \Join_{g\in L_N} \eval{g}{\Omega,G}^{\asbindings}.
\end{array}\]

In the following we define the semantics object constructs and with that $\eval{g}{\Omega,G}^{\asagraph}$ and $\eval{g}{\Omega,G}^{\asbindings}$.

%\vspace*{-1mm}
\mypar{\underline{Semantics of Object Construct Patterns}}
Let $\Omega$ be a set of bindings on $\gcgraph$ $G = (N, E, P, \rho, \delta, \lambda, \sigma)$.
Given an object construct $g$ with a grouping set $\grpvars$, we use $\grp{\Omega,g}$ to denote the grouping of $\Omega$ according to the grouping set $\grpvars$ of $g$.
The grouped set of bindings $\grp{\Omega,g}$ is the set of equivalence classes of $\Omega$ according to the equivalence relation $\sim_{\grpvars}$ such that $\mu_1\sim_{\grpvars}\mu_2$ if $\forall x\in\grpvars, \mu_1(x) = \mu_2(x)$ with $\mu_1,\mu_2 \in \Omega$.

In the following, we denote one equivalence class of $\op{grp}(\Gamma, g)$ as $\Omega'$ so that $\Omega'\in\op{grp}(\Gamma, g)$ holds.
$\Omega'$ is a set of bindings, where all $\mu\in\Omega'$ bind a grouping variable $x\in\grpvars$ to the same value.
We denote the value of a grouping variable $x$ in $\Omega'$ as $\Omega'(x)$.
Further, we denote the projection of $\Omega'$ to the grouping variables $\grpvars$ as $\Omega'(\grpvars)$ such that $\Omega'(\grpvars) = \set{x \mapsto \Omega'(x) \mid x\in\Gamma}$.

The semantics of an object construct $g$ with grouping set $\grpvars{}$ with respect to $\Omega$ evaluated on $G$ is the pair $\brk{\eval{g}{\Omega,G}^{\asagraph}, \eval{g}{\Omega,G}^{\asbindings}}$ where $\eval{g}{\Omega,G}^{\asagraph}$ is the $\gcgraph$ defined as
%\vspace*{-2mm}
$$\eval{g}{\Omega,G}^{\asagraph} = \bigcup_{\Omega' \in \grp{\Omega,g}} \eval{g}{\Omega',G}^{\asagraph}$$
and $\eval{g}{\Omega,G}^{\asbindings}$ is the set of bindings defined as
%\vspace*{-2mm}
$$\eval{g}{\Omega,G}^{\asbindings} = \bigcup_{\Omega' \in \grp{\Omega,g}} \eval{g}{\Omega',G}^{\asbindings}.$$

We next define the semantics of $\eval{g}{\Omega',G}^{\asagraph}$ and $\eval{g}{\Omega',G}^{\asbindings}$ for each object construct, assuming $G=(N, E, P, \rho, \delta, \lambda, \sigma)$.

\begin{itemize}
    \item If $g$ is a node construct \nodecpattern{x}{\grpvars{}}{S}, then 
        %\vspace*{-1mm}
        \begin{align*}
            \eval{g}{\Omega',G}^{\asagraph} =&\ (\set{v},\emptyset,\emptyset,\emptyset,\emptyset,
            %\restrict{\lambda}{x},\restrict{\sigma}{x})\\
            \eval{S}{x,v,\Omega',G}^{\aslabels},\eval{S}{x,v,\Omega',G}^{\asproperties}) \\
            \eval{g}{\Omega',G}^{\asbindings} =&\ \set{\set{x\mapsto v} \cup \Omega'(\grpvars{})}
        \end{align*}
        where
        %\vspace*{-2mm}
        %\begin{align*}
        %    \quad\quad
        $\quad v =\
        \begin{cases}
            \Omega'(x) & \text{if } x\in\boundvars \text{ and } \Omega'(x) \text{ is defined} \\
            \newobj{x,\Omega'(\grpvars{})} & \text{if } x\notin\boundvars.
        \end{cases} \\
        $%\end{align*}
        Here  \newobj{x,\Omega'(\grpvars{})} is a skolem function, returning a new distinct identifier for distinct values of $x$ and $\Omega'(\grpvars{})$.
        In case $x\in\boundvars$ and $\Omega'(x)$ is undefined, then $\eval{g}{\Omega',G}^{\asagraph}=\emptygraph$ and $\eval{g}{\Omega',G}^{\asbindings}=\emptybinding$.
        %Notice that $\Omega'(x)$ and $\Omega'(\grpvars{})$ are well defined, as all $\mu\in\Omega'$ have the same value on $\grpvars{}$ and on $x$, for $x\in\boundvars$, by definition of $\Omega'$.

    \item If $g$ is an edge construct \edgecpattern{x}{z}{\grpvars{}}{y}{S} and $\Omega'(x)$ and $\Omega'(y)$ are defined, then 
        %\vspace*{-1mm}
        \begin{align*}
            \eval{g}{\Omega',G}^{\asagraph} =&\ (\set{v,u},\set{e},\emptyset,\set{e\mapsto(v,u)},\emptyset,
              \eval{S}{z,e,\Omega',G}^{\aslabels},\eval{S}{z,e,\Omega',G}^{\asproperties}) \\
            \eval{g}{\Omega',G}^{\asbindings} =&\ \{\set{z\to e} \cup \Omega'(\grpvars{})\}
        \end{align*}
        where
        %\vspace*{-2mm}
        %\begin{align*} 
        %    \quad\quad 
        $\quad e =
        \begin{cases}
            \Omega'(z) & \text{if } z\in\boundvars \text{ and } \Omega'(z) \text{ is defined} \\
            \newobj{z,\Omega'(\grpvars{})} & \text{if } z\notin\boundvars
        \end{cases} \\
        $ %\end{align*}
        and $v=\Omega'(x)$ and $u=\Omega'(y)$.
        In case $\Omega'(x)$ or $\Omega'(y)$ is undefined, or in case $x\in\boundvars$ and $\Omega'(x)$ is undefined, then $\eval{g}{\Omega',G}^{\asagraph}=\emptygraph$ and $\eval{g}{\Omega',G}^{\asbindings}=\emptybinding$.
        This prevents dangling edges.

    \item If $g$ is a path construct \pathcpattern{x}{u}{@}{y}{S} and $\Omega'(u)$ is defined, then 
        %\vspace*{-2mm}
        \begin{align*}
            \eval{g}{\Omega',G}^{\asagraph} =&\ (N',E',\set{p},\set{e\mapsto\rho(e)\mid e\in\edges{p}},\set{p\mapsto\delta(p)},\lambda',\sigma') \\
            \eval{g}{\Omega',G}^{\asbindings} =&\ \{\Omega'(\set{x,y}) \cup \set{u\mapsto p}\}
        \end{align*}
        where $p=\Omega'(u)$ and
        %\vspace*{-1mm}
        \begin{align*}
            \lambda' =&\ \restrict{\lambda}{\nodes{u}\cup\edges{u}}\cup\eval{S}{u,p,\Omega',G}^{\aslabels} \\
            \sigma'  =&\ \restrict{\sigma}{\nodes{u}\cup\edges{u}}\cup\eval{S}{u,p,\Omega',G}^{\asproperties}
        \end{align*}
        If $\Omega'(u)$ is undefined, then $\eval{g}{\Omega',G}^{\asagraph}=\emptygraph$ and $\eval{g}{\Omega',G}^{\asbindings}=\emptybinding$.
        Note that $\Omega'(x)$ and $\Omega'(y)$ are defined, if $\Omega'(u)$ is defined, since $u$, $x$, and $y$ are all bound by the same path pattern in the \match{} clause. 

    \item If $g$ is a path construct \pathcpattern{x}{u}{}{y}{S} and $\Omega'(u)$ is defined, then 
        %\vspace*{-1mm}
        \begin{align*}
            \eval{g}{\Omega',G}^{\asagraph}   =&\ (\nodes{u},\edges{u},\emptyset,\set{e\mapsto\rho(e)\mid e\in\edges{p}}, \emptyset,\lambda',\sigma') \\
            \eval{g}{\Omega',G}^{\asbindings} =&\ \{\Omega'(\set{x,y})\}
        \end{align*}
        where 
        $\lambda' = \restrict{\lambda}{\nodes{u}\cup\edges{u}}$ and 
        $\sigma'  = \restrict{\sigma}{\nodes{u}\cup\edges{u}}$.
        If $\Omega'(u)$ is undefined, then $\eval{g}{\Omega',G}^{\asagraph}=\emptygraph$ and $\eval{g}{\Omega',G}^{\asbindings}=\emptybinding$.
\end{itemize}

%\vspace*{-1mm}
\mypar{\underline{Set and Remove Assignments}}
In the practical syntax, \setcl{} and \remove{} assignments can appear either inlined in the object pattern or in the \setcl--\remove{} sub-clause of the \construct{} clause.
In both cases, an assignment refers to a particular object construct by means of a construct variable.
In the following, we assume all assignments are inlined and appear in set of assignments $S$ of the respective object construct.

For a construct variable $x$, a variable $y\in\boundvars \cap (\cN \cup \cE \cup \cP)$, a label $l\in\bL$, a property key $k\in\bK$, and an expression $\expr$, assignments are syntactically defined as follows:
\begin{itemize}
    \item \copypattern{x}{y}, \setlabel{x}{l}, and \setprop{x}{k}{\expr} are \setcl{} assignments;
    \item \removelabel{x}{l}, and \removeprop{x}{k} are \remove{} assignments.
\end{itemize}

For a variable $x$ and an object identifier $o$, the semantics of $S$ is defined as 
$\eval{S}{x,o,\Omega',G}^{\aslabels}     = (\restrict{\lambda}{o}\cup\;\lambda_S)\setminus\lambda_R$
and
$\eval{S}{x,o,\Omega',G}^{\asproperties} = (\restrict{\sigma}{o} \cup\;\sigma_S)\setminus\sigma_R$
%\noindent
where 
$\restrict{\lambda}{o}$ is defined as expected and $\restrict{\sigma}{o}\ = \set{\brk{o,k}\mapsto v \mid \brk{o,k}\mapsto v \in \sigma \wedge \lnot\exists v', \brk{o,k}\mapsto v' \in \sigma_S}$, and
%\vspace*{-2mm}
\begin{align*}
    \lambda_S =   &\ \set{o\mapsto l \mid \Omega'(y)\mapsto l \in \lambda \wedge \copypattern{x}{y} \in S}  \\
              \cup&\ \set{o\mapsto l \mid \setlabel{x}{l}\in S} \\
    \lambda_R =   &\ \set{o\mapsto l \mid \removelabel{x}{l}\in S} \\
    \sigma_S  =   &\ \set{\brk{o,k}\mapsto v \mid \brk{\Omega'(y),k}\mapsto v \in \sigma \wedge \copypattern{x}{y} \in S} \\
              \cup&\ \set{\brk{o,k}\mapsto \eval{\expr}{\Omega', G} \mid \setprop{x}{k}{\expr}\in S} \\
    \sigma_R  =   &\ \set{\brk{o,k}\mapsto v \mid \brk{o,k}\mapsto v \in \sigma \wedge \removeprop{x}{k}\in S}.
\end{align*}
In essence, all \setcl{} assignments override values of existing properties $k$ and are applied before all \remove{} assignments, so that any expression $\expr$ used in a \setcl{} assignment sees all existing properties.

%\vspace*{-1mm}
\mypar{\underline{Full Construction Patterns}}
We now have the necessary ingredients to define the semantics of \construct{} clauses.
The syntax of such expressions is given by the following grammar:
%\vspace*{-1mm}
\begin{align*}
\cclause ::= &\ \csr{\fcp} \\ %\construct \ \fcp \\
\fcp     ::= &\ \bcp \ \mid \ \unioncpattern{\bcp}{\fcp} 
\end{align*}
%\mbox{\vspace*{-10mm} ... }
Note that we are not considering \setcl--\remove{} sub-clauses here.

Given a full construct $f$ consisting of a set of basic constructs $L$, we define the semantics of \construct{} clause as follow.
%\vspace*{-1mm}
%$$\eval{f}{\Omega,G} = \bigcup_{g\in L} \eval{g}{f,\Omega,G}^{\asagraph} .$$
\begin{align*}
    \eval{\csr{f}}{\Omega,G}   =&\ \bigcup_{b\in L} \eval{b}{\Omega,G}^{\asagraph}
\end{align*}
%\vspace*{-4mm}

% flatex input: [construct-example.tex]
%!TEX root = main.tex

\newcommand{\val}[1]{\ensuremath{\text{{\scriptsize{\bf #1}}}}}
\newcommand{\propk}[1]{\ensuremath{\textsf{\textbf{\scriptsize{#1}}}}}
\newcommand{\lbln}[1]{\ensuremath{\textsf{\textbf{\scriptsize{:#1}}}}}

\mypar{\underline{Example}} 
Consider a \csr{\conditionalcpattern{\set{f,g,h}}{\top}} as shown in Line~\ref{line:graphaggr} without \lstinline{social_graph}, where 
\begin{align*}
f =&\ \nodecpattern{x}{\set{e}}{\set{\setlabel{x}{\elt{Company}},\setprop{x}{\elt{name}}{e}}},\\
g =&\ \nodecpattern{n}{\set{n}}{\emptyset}, \text{ and }
h = \edgecpattern{n}{y}{\set{x,e,n}}{x}{\set{\setlabel{e}{\elt{worksAt}}}} .
\end{align*}

Let's assume the \match{} clause has the following set of bindings when evaluated on the \gcgraph{} $G$ in Figure \ref{fig:small-social-network}:
% \begin{center}
% $\Omega =$
% {\scriptsize\begin{tabular}{|@{\;}l@{\;}|@{\;}l@{\;}|}
% \hline
% {\bf n}        & {\bf e}\\
% \hline
% \hline
% {\bf \#Frank}  & {\bf "MIT"}\\
% {\bf \#Frank}  & {\bf "CWI"}\\
% {\bf \#Alice}  & {\bf "Acme"}\\
% {\bf \#Celine} & {\bf "HAL"}\\
% {\bf \#John}   & {\bf "Acme"}\\
% \hline 
% \end{tabular}} 
% \end{center}
\begin{align*}
\Omega =&\ \{\ \set{(n\mapsto\val{\#Frank}),(e\mapsto\val{\#MIT})},\set{(n\mapsto\val{\#Frank}),(e\mapsto\val{\#CWI})}, \\
        &\ \ \ \set{(n\mapsto\val{\#Alice}),(e\mapsto\val{\#Acme})},\set{(n\mapsto\val{\#Celine}),(e\mapsto\val{\#HAL})},\set{(n\mapsto\val{\#John}),(e\mapsto\val{\#Acme})} \ \}.
\end{align*}
We then have $\eval{\set{f,g,h}}{\Omega,G} = G_N \cup \eval{h}{\Omega_N,G}^{\asagraph}$ with
\begin{align*}
G_N      =&\ \eval{f}{\Omega,G}^{\asagraph} \cup \eval{g}{\Omega,G}^{\asagraph} \\
         =&\ (\set{\val{\#HAL},\val{\#Acme},\val{\#MIT},\val{\#CWI}},\emptyset,\emptyset,\emptyset,\emptyset,\quad\set{\val{\#HAL}\mapsto\lbln{Company},\ldots},\set{(\val{\#HAL},\propk{name})\mapsto\val{"HAL"},\ldots}) \\
      \cup&\ (\set{\val{\#John},\val{\#Frank},\val{\#Alice},\val{\#Celine}},\emptyset,\emptyset,\emptyset,\emptyset,\quad\set{\val{\#John}\mapsto\lbln{Person},\ldots},\set{(\val{\#John},\propk{fristname})\mapsto\val{"John"},\ldots}) \\
         =&\ (\set{\val{\#HAL},\val{\#Acme},\val{\#MIT},\val{\#CWI},\val{\#John},\val{\#Frank},\ldots},\ldots)\\
\Omega_N =&\ \eval{\nodecpattern{x}{\set{e}}{S_x}}{\Omega,G}^{\asbindings} \Join \eval{\nodecpattern{n}{\set{n}}{\emptyset}}{\Omega,G}^{\asbindings} \\
         =&\ \Omega \Join
\text{{\scriptsize\begin{tabular}{|@{\;}l@{\;}|@{\;}l@{\;}|}
\hline
{\bf x}        & {\bf e}\\
\hline
\hline
{\bf \#HAL}    & {\bf "HAL"}\\
{\bf \#Acme}   & {\bf "Acme"}\\
{\bf \#MIT}    & {\bf "MIT"}\\
{\bf \#CWI}    & {\bf "CWI"}\\
\hline 
\end{tabular}}}
\Join
\text{{\scriptsize\begin{tabular}{|@{\;}l@{\;}|}
\hline
{\bf n}        \\
\hline
\hline
{\bf \#Frank}  \\
{\bf \#Alice}  \\
{\bf \#Celine} \\
{\bf \#John}   \\
\hline 
\end{tabular}}}
=
\text{{\scriptsize\begin{tabular}{|@{\;}l@{\;}|@{\;}l@{\;}|@{\;}l@{\;}|}
\hline
{\bf n}        & {\bf x}        & {\bf e}\\
\hline
\hline
{\bf \#Frank}  & {\bf \#MIT}    & {\bf "MIT"}\\
{\bf \#Frank}  & {\bf \#CWI}    & {\bf "CWI"}\\
{\bf \#Alice}  & {\bf \#Acme}   & {\bf "Acme"}\\
{\bf \#Celine} & {\bf \#HAL}    & {\bf "HAL"}\\
{\bf \#John}   & {\bf \#Acme}   & {\bf "Acme"}\\
\hline 
\end{tabular}}}
\end{align*}
so that $\eval{h}{\Omega_N,G} = (N,E,\emptyset,\rho,\emptyset,\lambda,\emptyset)$ with
\begin{align*}
N       =&\set{\val{\#John},\val{\#Acme},\val{\#Frank},\val{\#MIT},\val{\#Frank},\val{\#CWI},\ldots},\\
E       =&\set{\val{\#1},\val{\#2},\val{\#3},\val{\#4},\val{\#5}},\\
\rho    =&\set{\val{\#1}\mapsto(\val{\#John},\val{\#Acme}),\val{\#2}\mapsto(\val{\#Frank},\val{\#MIT}),\val{\#2}\mapsto(\val{\#Frank},\val{\#CWI}),\ldots}, \text{ and}\\
\lambda =&\set{\val{\#1}\mapsto\lbln{worksAt},\ldots}.
%\eval{L}{\Omega,G} =&\ 
\end{align*} 
Finally, $\eval{\set{f,g,h}}{\Omega,G} = (N,E,\emptyset,\rho,\emptyset,\lambda,\sigma)$ with
\begin{align*}
%N       =&\set{\val{\#John},\val{\#Acme},\val{\#Frank},\val{\#MIT},\val{\#Frank},\val{\#CWI},\ldots},\\
%E       =&\set{\val{\#1},\val{\#2},\val{\#3},\val{\#4},\val{\#5}},\\
%\rho    =&\set{\val{\#1}\mapsto(\val{\#John},\val{\#Acme}),\val{\#2}\mapsto(\val{\#Frank},\val{\#MIT}),\val{\#2}\mapsto(\val{\#Frank},\val{\#CWI}),\ldots},\\
\lambda =&\set{\val{\#John}\mapsto\lbln{Person},\val{\#Acme}\mapsto\lbln{Company},\val{\#1}\mapsto\lbln{worksAt},\ldots}, \text{ and}\\
\sigma  =&\set{(\val{\#John},\propk{fristname})\mapsto\val{"John"},(\val{\#Acme},\propk{name})\mapsto\val{"Acme"},\ldots}.
%\eval{L}{\Omega,G} =&\ 
\end{align*}

% flatex input end: [construct-example.tex]

%\vspace*{-4mm}

% flatex input end: [construct.tex]

%\input{where}
% flatex input: [pathclause.tex]
%!TEX root = main.tex

%\vspace*{-4mm}
\subsection{The PATH clause }\label{sec:path}
A frequent idiom is to define a temporary view consisting of paths specified using a {\em path pattern},  optionally associating a weight to each path in the view.
The grammar for path clause expressions is:
\begin{align*}
    \bpclause  ::=&\ \Path \ \pname \  \kw{=} \ \ppattern  \\
    \pwcclause ::=&\ \bpclause \ \kw{Cost} \ \fexp \ \\%\As \ \cn \\
    \ppattern  ::=&\ \lbgp \ \mid \ \lbgp; \ \bgps \\
    \bgps ::=&\ \bgp \ \mid \ \bgp, \ \bgps \\
    \lbgp ::=&\ \ep \ \mid \ \ep, \ \lbgp \ \mid \\ 
             &\ \ppattern \ \mid \ \ppattern, \ \lbgp
\end{align*}
where $\pname$ is a path view name and $\fexp$ is a cost function expression.
%$\fexp$ is a cost function expression, and $\cn$ is a property name (that is, $\cn \in \bK$).  
Note that $\ppattern$ can refer to path views defined by other \Path{} clauses appearing before it in the head clause where it is defined.

In the syntax, we additionally require that walk patterns are joinable.  
%We say a graph pattern $\varphi$ is linear if it conforms to the following grammar: (1) $\varphi$ is a basic graph pattern; or, (2) $\varphi = b, \varphi'$, where $b$ is a basic graph pattern, $\varphi'$ is a linear graph pattern, and $b$ and $\varphi'$ are joinable.
Given a walk pattern $b_1, \ldots, b_n$, where $n \geq 1$ and each $b_i$ is either an edge pattern or a path pattern, we say that $b_1, \ldots, b_n$ is joinable if for every $i \in \{1, \ldots, n-1\}$:
\begin{itemize}
    \item $b_i$ is either an edge pattern $\edgepattern{x}{~z~}{y}$ or a path pattern $\pathpattern{x}{w}{r}{y}$ or a path pattern $\wpathpattern{x}{w}{r}{y}$,  $b_{i+1}$ is either an edge pattern $\edgepattern{x'}{~z'~}{y'}$ or a path pattern $\pathpattern{x'}{w'}{r'}{y'}$ or a path pattern $\wpathpattern{x'}{w'}{r'}{y'}$, and $y = x'$.
\end{itemize}
%\begin{itemize}
%    \item $b$ is a node pattern $(x)$, $b_1$ is an edge pattern 
%        $\edgepattern{x'}{~z~}{y}$, and $x=x'$ (and likewise if $b_1$ is a node or path pattern), or, 
%    \item $b$ is an edge pattern $\edgepattern{x}{~z~}{y}$,
%        $b_1$ is an edge pattern $\edgepattern{x'}{~z'~}{y'}$,
%        and $y = x'$ (and likewise for other combinations of edge and/or path patterns).
%\end{itemize}
The interpretation of a walk pattern in a graph $G$ is a path in $G$.

\noindent We express path views via a \Path{} clause, defined as follows:
\begin{align*}
\pclause ::=  \, & \bpclause \  \mid \ \pwcclause \  \mid \\
& \bpclause  \ \where \ \bc \ \mid\\
& \pwcclause  \ \where \ \bc 
\end{align*}

\noindent Intuitively, the evaluation of \\
%\vspace*{-2mm}
%\begin{multline*}
$$\Path \ \pname \  \kw{=} \ \ppattern \ %\\
\cost \ \fexp \ %\As \ \cn \ 
                \where \ \bc$$
%\end{multline*}
on a $\gcgraph$ $G$ is a view $\pname$ consisting of a set of paths, each satisfying $\bc$.  
The paths are given by bindings of $\ppattern = \lbgp; \ \bgps$ in $G$. More precisely, assuming that $x, y \in \cN$ are the starting and ending nodes of $\lbgp$, for every pair of nodes $a$, $b$ in $G$, we look for the shortest path $L$ from $a$ to $b$ over $G$ that conforms to $\lbgp$ and satisfies the other graph patterns in $\ppattern$, and if such a path exists then we check that it satisfies $\bc$. If this holds, then we create a binding $\mu$ and a fresh path identifier $p$ such that $\dom{\mu} = \{x, y, \pname\}$, $\mu(x) = a$, $\mu(y) = b$, $\mu(\pname) = p$ and $p$ is associated to $L$. 
%$\mu$ representing a shortest path conforming to $\lbgp$ and satisfying the other graph expressions occurring in $\ppattern$, and then we add $\mu$ to $\pname$.
Finally, the evaluation of the path clause is a set containing all such bindings $\mu$.
%Finally, each path identifier in the view holds in property $\cn$ the value given by cost function $\fexp$ on the path.
It is straightforward to see how these ideas can be formalized by using the notation considered previously.
%in the previous sections. 

Regarding the use of path views,
recall that \pname{} is defined in the head clause of a full graph query $\varphi$.
For any path pattern appearing in $\varphi$, 
\pname{} can be used as a symbol in the regular expression over which the pattern is evaluated.
\pname{} is then prefixed with the tilde symbol ``\textasciitilde'' to indicate that reference is being made to the binary view $\pname(x, y)$ (rather than an element of \bL).

% flatex input end: [pathclause.tex]

%\input{where}
% flatex input: [fullquery.tex]
%!TEX root = main.tex
%\vspace*{-1mm}
\subsection{Basic and Full Graph Queries}\label{sec:fullquery}

Recall that a basic graph query is defined as a CONSTRUCT clause followed by a MATCH clause, i.e., a basic graph query has the form
%\begin{eqnarray*}
$\kw{Construct}\ \Phi\ \kw{Match}\ \Psi$.
%\end{eqnarray*}
The evaluation of such a query over a graph $G$ is defined as:
%\begin{eqnarray*}
$\eval{\kw{Construct}\ \Phi\ \kw{Match}\ \Psi}{G}  =  \eval{\Phi}{\Omega,G}$,
%\end{eqnarray*}
where $\Omega = \eval{\Psi}{\Omega',G}$ and $\Omega'$ is either the set of bindings provided by an outer query if $\Psi$ is an inner query or a singleton set containing the binding $\mu_\emptyset$ with empty domain (recall that $\mu_\emptyset$ is compatible with every other binding).
%Given the semantics of the construct clause, a basic query $q$ consisting of a match clause $m$ and a construct clause $c$ evaluates on a $\gcgraph$ $G$ to another $\gcgraph$ $G'$ defines as follows:
%$$G' = \eval{q}{G} = \eval{c}{\Omega,G} \ \text{with} \ \Omega = \eval{m}{\Omega_O,G},$$
%where $\Omega_O$ is either the set of mappings provided by the outer query if $q$ an inner query or $\set{\emptyset}$ if $q$ is the outer most query.
%As can be seen here, basic query are the scope of variable bindings.
%We defined the semantics of basic graph queries. 
A full graph query is obtained by using the operators \Union, \Intersect{}, and \Minus{} to combine basic graph queries. Thus, to complete the definition of full graph queries, we need to provide the semantics of these operators on \gcgraph's. 
%=======
%So far we have defined the semantics of basic graph queries. A full graph query is obtained by using the operators \Union, \Intersect{}, and \Minus{} to combine basic graph queries. 
%We define full graph queries by providing the semantics of these operators on \gcgraph's. 
%>>>>>>> c40a0e4dd4690e038cf575f81f8e09d6875bc0da

%Next we define the semantics of full queries. 
%Let $G_1 = (N_1, E_1, P_1, \rho_1, \delta_1, \lambda_1, \sigma_1)$
%and $G_2 = (N_2, E_2, P_2, \rho_2, \delta_2, \lambda_2, \sigma_2)$ be $\gcgraph$.  We say $G_1$ and $G_2$ are {\em consistent} if
%\begin{itemize}
%\item For every $e \in E_1\cap E_2$, it holds that $\rho_1(e) = \rho_2(e)$.
%\item For every $p \in P_1\cap P_2$, it holds that $\delta_1(e) = \delta_2(e)$.
%\item For every $o\in ((N_1\cap N_2)\cup (E_1\cap E_2)\cup (P_1\cap P_2))$, it holds that
%$\lambda_1(o) = \lambda_2(o)$.
%\item For every $o\in ((N_1\cap N_2)\cup (E_1\cap E_2)\cup (P_1\cap P_2))$ and $k\in\bK$, it holds that $\sigma_1(o, k) = \sigma_2(o, k)$.
%\end{itemize}

In what follows, assume that $G_1 = (N_1, E_1, P_1, \rho_1, \delta_1, \lambda_1, \sigma_1)$ and $G_2 = (N_2, E_2, P_2, \rho_2, \delta_2, \lambda_2, \sigma_2)$ are $\gcgraph$.  Moreover, say that $G_1$ and $G_2$ are consistent if: 
%\begin{itemize}
%\item 
(i) for every $e \in E_1\cap E_2$, it holds that $\rho_1(e) = \rho_2(e)$, and
%\item 
(ii) for every $p \in P_1\cap P_2$, it holds that $\delta_1(p) = \delta_2(p)$.
%\end{itemize}

%\vspace*{-1mm}
\mypar{\underline{The union operator}} If $G_1$ and $G_2$ are not consistent, then $G_1 \cup G_2$ is defined as the empty $\gcgraph$. Otherwise, 
%\begin{eqnarray*}
$G_1 \cup G_2  =  (N_1 \cup N_2, E_1 \cup E_2, P_1 \cup P_2, \rho, \delta, \lambda, \sigma)$,
%\end{eqnarray*}
where (i)
%\begin{itemize}
%\item 
for every $x \in (N_1 \cup N_2 \cup E_1 \cup E_2 \cup P_1 \cup P_2)$ and $k \in \bK$: $\lambda(x)\!=\!\lambda_1(x) \cup \lambda_2(x)$ and $\sigma(x,k)\!=\!\sigma_1(x,k) \cup \sigma_2(x,k)$;
%\item 
(ii) for every $e \in E_1 \cup E_2$:
%\vspace*{-2mm}
\begin{eqnarray*}
\rho(e) & = & 
\begin{cases}
\rho_1(e) & \text{if } e \in E_1\\
\rho_2(e) & \text{otherwise;}
\end{cases}
\end{eqnarray*}
%\vspace*{-2mm}
%\item 
and (iii) for every $p \in P_1 \cup P_2$:
%\vspace*{-2mm}
\begin{eqnarray*}
\delta(p) & = & 
\begin{cases}
\delta_1(p) & \text{if } p \in P_1\\
\delta_2(p) & \text{otherwise.}
\end{cases}
%\vspace*{-2mm}
\end{eqnarray*}
%\end{itemize}
%=======
%where:
%\begin{itemize}
%\item For every $x \in (N_1 \cup N_2 \cup E_1 \cup E_2 \cup P_1 \cup P_2)$ and $k \in \bK$, it holds: $\lambda(x)\!=\!\lambda_1(x) \cup \lambda_2(x)$ and $\sigma(x,k)\!=\!\sigma_1(x,k) \cup \sigma_2(x,k)$.
%
%\item For every $e \in E_1 \cup E_2$:
%%\vspace*{-2mm}
%%\begin{eqnarray*}
%$\rho(e)  = 
%\begin{cases}
%\rho_1(e) & \text{if } e \in E_1\\
%\rho_2(e) & \text{otherwise.}
%\end{cases}$
%%\end{eqnarray*}
%%\vspace*{-2mm}
%
%
%\item For every $p \in P_1 \cup P_2$:
%% \vspace*{-2mm}
%% \begin{eqnarray*}
%$\delta(p)  = 
%\begin{cases}
%\delta_1(p) & \text{if } p \in P_1\\
%\delta_2(p) & \text{otherwise.}
%\end{cases}$
%% \vspace*{-2mm}
%% \end{eqnarray*}
%\end{itemize}
%
%>>>>>>> c40a0e4dd4690e038cf575f81f8e09d6875bc0da

%\vspace*{-1mm}
\mypar{\underline{The intersection operator}} If $G_1$ and $G_2$ are not consistent, then $G_1 \cap G_2$ is defined as the empty $\gcgraph$. Otherwise, 
%\begin{eqnarray*}
$G_1 \cap G_2  =  (N_1 \cap N_2, E_1 \cap E_2, P_1 \cap P_2, \rho, \delta, \lambda, \sigma)$,
%\end{eqnarray*}
where
%\begin{itemize}
%\item 
(i) for every $x \in (N_1 \cap N_2) \cup (E_1 \cap E_2) \cup (P_1 \cap P_2)$ and $k \in \bK$: $\lambda(x)\!=\!\lambda_1(x) \cap \lambda_2(x)$ and $\sigma(x,k)\!=\!\sigma_1(x,k) \cap \sigma_2(x,k)$;
%\item 
(ii) for every $e \in E_1 \cap E_2$: $\rho(e) = \rho_1(e)$; and
%\item 
(iii) for every $p \in P_1 \cap P_2$: $\delta(p) = \delta_1(p)$.
%\end{itemize}

%\vspace*{-1mm}
\mypar{\underline{The difference operator}} The difference of $G_1$ and $G_2$ is defined~as:
%\begin{eqnarray*}
$G_1 \smallsetminus G_2  =  (N, E, P, \rho, \delta, \lambda, \sigma)$,
%\end{eqnarray*}
where:
%\begin{align*}
$N = N_1 \smallsetminus N_2$, 
$E = \{ e\in E_1 \smallsetminus E_2 \mid \rho_1(e)  =  (a, b), a\in N \text{ and } b\in N\}$,
$P =  \{ p\in P_1 \smallsetminus P_2 \mid \nodes{p} \subseteq N \text{ and } \edges{p} \subseteq E\}$, 
% P = & \{ p\in P_1 \smallsetminus P_2 \mid \delta_1(p) = [a_1, e_1, a_2, \ldots, a_n, e_n, a_{n+1}],\\
% & \hspace{55pt} a_i \in N \text{ for every } i \in \{1, \ldots, n+1\} \text{ and }\\
% & \hspace{55pt} e_j \in E \text{ for every } j \in \{1, \ldots, n\}\},
%\end{align*}
%and 
$\rho  =  \restrict{\rho_1}{E}$, $\delta  =  \restrict{\delta_1}{P}$, $\lambda  =  \restrict{\lambda_1}{(N\cup E \cup P)}$ and $\sigma = \restrict{\sigma_1}{(N\cup E \cup P)\times \bK}$.

\ignore{% moved to graph.tex
\subsection{The GRAPH clause and GRAPH VIEW}
To conclude with the formal definition of the semantics of \gcore, we need to
define the semantics of the \Graph{} clause and \Graph{} \View{}.
The \Graph{} \View{} clause is used to create permanent views that may be re-used by
multiple queries while the \Graph{} clause is used for the declaration of
query-local views only (similar to the \Path{} clause).
Both clauses are given by the following grammar:
\begin{eqnarray*}
\gclause & ::=  & \kw{Graph}\ \gid\ \kw{As}\ (\fgquery),\\
\gv & ::=  & \kw{Graph View}\ \gid\ \kw{As}\ (\fgquery).
\end{eqnarray*}
%where $\gid$ is a graph identifier. 
Assuming that $\Psi$ is a full graph query, the evaluation of clause $\kw{Graph}\ \gid\ \kw{As}\ (\Psi)$ over a graph $G$ associates the graph $\eval{\Psi}{G}$ to the identifier $\gid$, that is, $\graph(\gid) = \eval{\Psi}{G}$. The semantics of $\kw{Graph View}\ \gid\ \kw{As}\ (\Psi)$ is defined exactly in the same~way.
}
% flatex input end: [fullquery.tex]

%\input{where}
% flatex input: [graph.tex]
%\ignore{% sorry not enough space
\subsection{The GRAPH clause and GRAPH VIEW}
%\vspace*{-1mm}
%To conclude with the formal definition of the semantics of \gcore, we need to
%define the semantics of the \Graph{} clause and \Graph{} \View{}.
The \Graph{} \View{} clause is used to create permanent views that may be re-used by
multiple queries while the GRAPH clause is used for the declaration of
query-local views only (similar to the \Path{} clause).
Both clauses are given by the following grammar:
\begin{eqnarray*}
\gclause & ::=  & \kw{Graph}\ \gid\ \kw{As}\ (\fgquery),\\
\gv & ::=  & \kw{Graph View}\ \gid\ \kw{As}\ (\fgquery),
\end{eqnarray*}
%where $\gid$ is a graph identifier. 
Assuming that $\Psi$ is a full graph query, the evaluation of clause $\kw{Graph}\ \gid\ \kw{As}\ (\Psi)$ over a graph $G$ associates the graph $\eval{\Psi}{G}$ to the graph identifier $\gid$, that is, $\graph(\gid) = \eval{\Psi}{G}$. The semantics of $\kw{Graph View}\ \gid\ \kw{As}\ (\Psi)$ is defined exactly in the same~way.
%}

% flatex input end: [graph.tex]

%\input{where}

% flatex input end: [theoretical-results.tex]

%\clearpage
%\clearpage
%\input{complex_queries}
%\clearpage
%\input{interactive_queries}
%\clearpage
%\input{business_intelligence}

\end{document}